\documentclass[journal]{IEEEtran}
\IEEEoverridecommandlockouts
\usepackage{makecell}
\usepackage{cite}
\usepackage{amsmath,amssymb,amsfonts}

\usepackage{graphicx}
\usepackage{textcomp}
\usepackage{xcolor}
\newtheorem{theorem}{Theorem}

\usepackage{bbm}
\usepackage{bm}
\newtheorem{proof}{Proof}
\newtheorem{corollary}{Corollary}
\usepackage{bbm}
\usepackage{pifont}
\usepackage[table]{xcolor}
\usepackage{svg}

\usepackage{algorithm}
\usepackage{algpseudocode}

\newcounter{proposition}
\newenvironment{proposition}[1][]
  {\refstepcounter{proposition}%
   \par\indent\textit{Proposition~\theproposition
   \ifx\relax#1\relax\else~\cite{#1}\fi:}\ \ignorespaces}
  {\par}

\def\BibTeX{{\rm B\kern-.05em{\sc i\kern-.025em b}\kern-.08em
   T\kern-.1667em\lower.7ex\hbox{E}\kern-.125emX}}

\begin{document}
\setlength{\textfloatsep}{0.11cm}
\setlength{\baselineskip}{0.42cm}
\setlength{\abovecaptionskip}{1mm}

\title{Achievable Information Rate for FTN under a Reduced Complexity BCJR Receiver}
\author{Zichao~Zhang,~\IEEEmembership{Student Member,~IEEE,}
  Gokhan M. Guvensen, Melda~Yuksel,~\IEEEmembership{Senior Member,~IEEE,} 
Halim~Yanikomeroglu,~\IEEEmembership{Fellow,~IEEE}
\thanks{This work was funded in part by the Scientific and Technological Research Council of Turkey, TUBITAK BILGEM, and in part by a Discovery Grant awarded by the Natural Sciences and Engineering Research Council of Canada (NSERC).}
\thanks{Z. Zhang and H. Yanikomeroglu are with the Department of Systems and Computer Engineering at Carleton University, Ottawa, ON, K1S 5B6, Canada e-mail:	zichaozhang@cmail.carleton.ca, halim@sce.carleton.ca.}
\thanks{G. Guvensen and M. Yuksel are with the Department of Electrical and Electronics Engineering, Middle East Technical University, Ankara, 06800, Turkey, e-mail: \{ymelda, guvensen\}@metu.edu.tr.}
}

\maketitle

\begin{abstract}
    % This paper studies the achievable information rate (AIR) for both full-state and reduced-state Bahl–Cocke–Jelinek–Raviv (BCJR)  receiver with bidirectional decision feedback for faster-than-Nyquist (FTN) detection. 
    The capacity of faster-than-Nyquist (FTN) signaling in additive white Gaussian noise (AWGN) channels has been established and is achieved by Gaussian inputs. However, the capacity of FTN signaling with finite-constellations is not yet fully characterized. The achievable information rate (AIR) provides a lower bound on this constrained capacity by quantifying the rate supported by a specified input distribution and receiver. This paper investigates the AIR of FTN signaling with finite constellations using full-state and reduced-state Bahl–Cocke–Jelinek–Raviv (BCJR) receivers, with bidirectional decision feedback incorporated into the reduced-state receiver.
    Especially, the detection problem is formulated in a reproducing kernel Hilbert space (RKHS), where a geometric foundation is established to understand FTN under severe time packing.  The formulation establishes that the Ungerboeck metric remains valid even when the noise covariance is singular. 
    % Nyquist signaling with the same constellation reaches the same limiting AIR per symbol, but FTN achieves a higher spectral efficiency in this limit by transmitting symbols at a higher rate within the same bandwidth.
    We further characterize detection errors geometrically using RKHS distances between competing sequences.  We also study outage probability for reduced-state detection, which occurs when AIR is zero, in addition to deriving the zero-AIR outage criterion for reduced-state detection. Unlike ISI-free Nyquist signaling, reduced-state FTN detection can retain residual ISI that causes the receiver to assign excessive confidence to incorrect symbols, leading to outage even at high signal-to-noise ratio (SNR). Our results confirm both the spectral efficiency gains and the feasibility of FTN systems employing finite constellations and realistic receiver architectures. 
\end{abstract}

\section{Introduction}

The continued growth of wireless devices and data-intensive services is placing increasing pressure on the available time--frequency resources. Early vision documents for sixth-generation (6G) wireless systems identify immersive communication, global broadband, ubiquitous internet of things, spatio-temporal services, and critical services as important future use cases \cite{whatshould6Gbe}. 
%Therefore, there is an urgent need of bandwidth to comply with future communication demands. In order to increase transmission rate, there are several ways of accomplishing it. Higher frequency bands can be exploited to allow for more bandwidth, we can also increase spectral efficiency to allow more transmission data under the same bandwidth. 
Meeting the growing communication demands of these emerging services requires higher data rates. 
Faster-than-Nyquist (FTN) signaling addresses this objective by increasing the spectral efficiency while retaining the pulse shape and the occupied bandwidth \cite{mazo}.

 FTN signaling uses a symbol interval $\delta T$, where $T$ is the Nyquist interval and $\delta\in(0,1]$ is the acceleration factor. For $\delta<1$, the symbol rate increases by a factor of $1/\delta$ without changing the pulse shape or the occupied bandwidth, at the cost of intentional intersymbol interference (ISI). The capacity of multiple-input multiple-output (MIMO) FTN signaling has been characterized, with Gaussian inputs and appropriate precoding and power allocation achieving capacity \cite{zhang2022faster,zhang2025iapr}.  Practical systems; however, employ finite constellations such as QPSK or QAM, whose achievable rates are not necessarily the same as those achieved with Gaussian constellations. Constrained-capacity analysis for precoded FTN has therefore been investigated under finite-alphabet inputs \cite{zhang2026constrainedcapacityanalysisfasterthannyquist}. With practical receivers, the attainable rate further depends on how effectively the intentional ISI is handled. The achievable information rate (AIR) quantifies the rate supported by a specified input distribution and decoding metric, providing a lower bound on the corresponding constrained capacity. AIR has been used, for example, to evaluate reduced-complexity Ungerboeck receivers for quantized wideband massive single-carrier (SC)-MIMO \cite{Gokhanmismatch}. This motivates its use here to investigate how FTN acceleration and receiver complexity affect achievable rates.

% FTN transceivers can be broadly divided into frequency-domain and time-domain architectures \cite{1_R1}. Frequency-domain transceivers precode the symbols and use a transform or eigenchannel decomposition to obtain parallel ISI-free transmission, followed by symbol-wise processing. In time-domain designs, the information symbols are modulated directly on time-packed pulses and must be recovered from samples that contain ISI and correlated noise. This paper focuses on the time-domain architecture, where the receiver must resolve the intentional ISI, and more aggressive time packing increases the effective ISI span and receiver complexity.

The intentional ISI introduced by FTN motivates the use of receivers originally developed for ISI channels. Maximum-likelihood sequence estimation (MLSE) is a standard approach for such channels. Forney introduced a trellis-based MLSE receiver that uses a noise-whitening filter \cite{Forneyisi}, whereas Ungerboeck developed an observation-model receiver that operates directly on correlated samples \cite{ungerboeckfirstpaper}. The equivalence of these observation models was established in \cite{forneyungerequiv}.
These two approaches both utilize finite state machines, where a trellis is built, the likelihood of sequences is computed iteratively, and an optimal state path is determined. Both of the algorithms make decisions using the maximum likelihood (ML) principle. Meanwhile, the Bahl-Cocke-Jelinek-Raviv (BCJR) algorithm applies the maximum a posteriori (MAP) principle for making decisions, where the likelihood is computed using the sum-product algorithm  \cite{bcjr}. 
%All three methods require a trellis for recursive computation. For full-state detection, the trellis state must retain the relevant interfering-symbol history, whose length is determined by the channel’s effective delay spread, hereafter referred to as the ISI depth. However, under severe ISI, in order to cover the ISI depth, the state length also has to grow. As a result, the number of total states grows exponentially. If the transmitted symbols are drawn from an alphabet of size $M$, the state length is $L$, then the total number of states is $M^L$, as the state length grows the number of total states can easily become prohibitive. Therefore, in order to reduce the state length and achieve a smaller trellis size, reduced-state detection is needed.
All three methods rely on a trellis for recursive computation. In full-state detection,  the required state length increases rapidly. For an alphabet of size $M$ and state length $L$, the trellis contains $M^L$ states, which can become prohibitive. Reduced-state detection is therefore required to limit the trellis complexity.

Reduced-state receivers for ISI channels include decision-feedback Forney-type MLSE \cite{Forneyred}, bias-corrected Ungerboeck detection \cite{starkpaper}, and BCJR-type detection with bidirectional decision feedback \cite{Gokhanhocampaper}. Pulse shortening and state merging provide further complexity reductions \cite{svensson1984,svensson2002thesis}. For FTN, reduced-complexity approaches include M-BCJR receivers \cite{redcomplxftn,redcomplxftnunger}, successive symbol-by-symbol detection with go-back-$K$ \cite{lowcomplxsssse}, and channel-shortening-based MLSE \cite{mlseequchnlshort}.
Recent FTN receiver studies have extended reduced-complexity detection
beyond the trellis methods discussed above. In \cite{gokhanrecommend}, the authors investigated FTN spectral efficiency under finite
trellis-state and turbo-iteration constraints using LDPC-coded
transmission and EXIT-chart analysis. Che and Bai \cite{cheandbai} combine channel
shortening with an Ungerboeck M-BCJR detector. These methods reduce
the trellis complexity, but their performance depends on the selected
shortened response or code structure. 
However, the AIR supported by FTN with the bidirectional reduced-state
receiver of \cite{Gokhanhocampaper} remains insufficiently characterized. 
This receiver
is particularly suitable for investigating achievable rates because its
state length offers a controllable trade-off between complexity and
detection accuracy. Under full-state configuration, the receiver
approaches ML detection. We therefore use
this receiver to investigate how time packing and receiver complexity
affect the AIR of FTN signaling. We therefore use this receiver to investigate how time packing and receiver complexity affect the AIR of FTN signaling. 
The contributions of this paper are summarized as follows.
\begin{itemize}
    \item We analyze the AIR of FTN signaling with finite constellations using full-state BCJR detection and the practical reduced-state Ungerboeck-type receiver with bidirectional decision feedback proposed in \cite{Gokhanhocampaper}.
    \item We establish that the AIR with the full-state receiver approaches the constellation entropy at high signal-to-noise ratio (SNR), provided that the minimum RKHS distance between distinct transmitted sequences is positive. Under this condition, FTN and Nyquist signaling attain the same limiting information rate per symbol, while FTN achieves higher spectral efficiency through its increased symbol rate.
    \item We derive an exact decomposition of the reduced-state AIR into penalties associated with wrong-symbol score advantages and residual soft ambiguity. This decomposition yields a zero-AIR outage criterion and explains how persistent confidence in incorrect symbols can cause outage even at high SNR, limiting the rate gains from stronger time packing.
    \item 
     We establish the likelihood foundation for this analysis under singular matched-filter noise. Using an RKHS formulation, we prove that the noiseless signal and noise share the same supported subspace and then the Ungerboeck metric remains valid when the noise covariance is rank deficient.
    \item We provide a geometric characterization of reduced-state error events through the RKHS distance between competing sequences and the signed projections of residual future-symbol interference and noise. This characterization explains how state reduction and imperfect feedback affect the reliability of the decisions.
    % \item We derive the AIR of the reduced-state receiver directly from its final soft-output scores. The AIR is decomposed into a wrong-symbol score-advantage penalty and a residual soft-ambiguity term. Based on this decomposition, we obtain an exact outage criterion and show that persistent wrong-symbol score advantages can dominate at high SNR and drive the clipped AIR into outage.
\end{itemize}

The following symbols are used throughout this paper. The superscripts $*$, $T$, and $\dagger$ denote complex conjugate, transpose, and conjugate transpose, respectively. The symbol $\star$ denotes convolution and $\mathcal{R}(\bm{G})$ denotes
    the range of matrix ${\bm{G}}$. The notation $(\cdot)_{m,n}$ represents the $(m,n)$th entry of a matrix.  The terms $y(n\delta T)$ and $y[n]$ are used interchangeably to denote the $n$th time-domain sample of the continuous-time signal $y(t)$.  The operators $\mathrm{tr}(\cdot)$,  $\mathbb{E}[\cdot]$, and $\operatorname{var} (\cdot)$ denote the trace of a matrix, the expectation, and the variance, respectively. The operations $\doteq$, $\Longleftrightarrow$, and $\propto$ denote
equality up to a normalization constant, equivalence, and proportionality, respectively.    Finally, $\bm{I}_N$ represents the $N\times N$ identity matrix. For any indexed quantities $x_i$, the notation
$\{x_i\}_{i=m}^{n}$ denotes the ordered sequence
$(x_m,x_{m+1},\ldots,x_n)$. 
For clarity, we distinguish the transmitted sequence from candidate paths and receiver-generated feedback variables as follows:
\begin{itemize}
\item $\bar{\bm a}$ denotes the actual transmitted symbol sequence,
\item $\bm a$ denotes an arbitrary candidate sequence,
\item $\bar a_k$ denotes the actual transmitted symbol at time $k$,
\item $a_k$ denotes the symbol at time $k$ on a candidate trellis path,
\item $\hat a_k$ denotes the tentative future-symbol decision supplied by the backward recursion, and 
\item $\dot a_k$ denotes the traceback estimate on a surviving path.
\end{itemize}

The remainder of this paper is organized as follows. Section II introduces the FTN system model. Section III presents the reduced-state Ungerboeck-type receiver with bidirectional decision feedback. Section IV establishes the validity of the Ungerboeck metric in the presence of singular noise using a RKHS formulation. Section V provides a geometric analysis of pairwise error events, and Section VI investigates the achievable information rate (AIR) and its outage behavior. Section VII presents the simulation results, followed by the conclusions in Section VIII.

\section{System Model}

% Assume that a block of $N$ complex-valued symbols is transmitted over a complex additive white Gaussian noise (AWGN) channel. Each symbol is drawn independently and uniformly from a discrete alphabet $\mathcal{A}$ of size $M$, where $\mathcal{A}$ is zero mean. The average symbol energy is denoted by $E_s$. The transmitted symbols $\bar a_n$, $n=0,\dots,N-1$, are i.i.d. and uniformly distributed over $\mathcal{A}$, and satisfy
% \begin{equation}
%     \mathbb{E}[\bar a_n]=0,
%     \qquad
%     \mathbb{E}[|\bar a_n|^2]=E_s.
% \end{equation}

Assume that a block of $N$ complex-valued symbols is transmitted, where $\mathcal{A}$ denotes a discrete alphabet containing $M$ complex symbols and the data symbols $\bar a_n$, $n=0,\ldots,N-1$, are independent and identically distributed (i.i.d), such that
\begin{equation}
    \mathbb{E}[\bar a_n]=0,
    \qquad
    \mathbb{E}[|\bar a_n|^2]=E_s.
\end{equation}
The transmitted data vector $\bar{\bm a}\in\mathcal{A}^N$ is defined as
\[
    \bar{\bm a}=[\bar a_0,\ldots,\bar a_{N-1}]^T.
\]

The data symbols are modulated using a unit-energy pulse $p(t)$ for pulse shaping, where
\begin{equation}
    \int_{-\infty}^{\infty}|p(t)|^2\,\mathrm{d}t=1.
\end{equation}
A root-raised-cosine (RRC) or square-root-raised-cosine (SRRC) pulse with roll-off factor $\beta$ are two possible choices; however, the subsequent derivations do not depend on a particular pulse shape. The symbols are transmitted every $\delta T$ seconds. For a sinc pulse, $T=\frac{1}{W}$, with $W$ denoting the occupied bandwidth. For an SRRC pulse, the occupied bandwidth is $(1+\beta)W$.  The transmitted waveform is therefore
\begin{equation}
    x(t)=\sum_{n=1}^{N}\bar a_n
    p\!\left(t-n\delta T\right).
    \label{eqn:tx_waveform}
\end{equation}
The physical channel is assumed to be a circularly symmetric complex Gaussian noise channel. The noise term is denoted with $w(t)$ and the noise power spectral density is $N_0$, i.e,
\[
    \mathbb{E}[w(t)w(s)]=N_0\delta_D(t-s),
\]
where $\delta_D(\cdot)$ denotes the Dirac delta function. The receiver applies the matched filter $p^*(-t)$ to maximize the output SNR. The received  waveform after matched filtering is characterized by the pulse-correlation function
    $g(t)\triangleq p(t)\star p^*(-t)$.
The noise term after the matched filter is denoted as
    $\eta(t)\triangleq w(t)\star p^*(-t)$.
Consequently, the matched-filter output is
\begin{equation}
    y(t)=\sum_{n=1}^{N}\bar a_n
    g\!\left(t-n\delta T\right)+\eta(t).
    \label{eqn:mf_output}
\end{equation}
This matched-filter output is sampled every $\delta T$ seconds. Defining  $g_i=g(i\delta T)$, $y_i=y(i\delta T)$, and $\eta_i=\eta(i\delta T)$, and collecting the samples into vectors $\bm y=[y_0,\ldots,y_{N-1}]^T$ and  $\bm\eta=[\eta_0,\ldots,\eta_{N-1}]^T$, the discrete-time input--output relationship of the FTN system can be written as
\begin{align}
    \bm y=\bm G\bar{\bm a}+\bm\eta, \label{eqn:sysmodelunger}
\end{align}
where $\bm{G}$ is defined by $(\bm{G})_{i,j}=g_{i-j}$ for $i,j=0,\dots,N-1$.
 Since the entries of $\bm{G}$ depend only on the difference between the row and column indices, $\bm{G}$ is a Toeplitz matrix \cite{gray}. It is also Hermitian because $g(t)$ is the pulse-correlation function. Therefore, $\bm{G}$ describes the ISI coupling among the transmitted symbols. Under Nyquist signaling, $\bm{G}$ becomes an identity matrix, whereas under FTN signaling, the off-diagonal entries of $\bm{G}$ are generally non-zero.

We assume that the energy of the composite pulse $g(t)$ is approximately confined to the interval $(-L\delta T,L\delta T)$. Accordingly,
\[
    g_l=0,\qquad \forall |l|>L.
\]
Under this finite-memory assumption, $\bm{G}$ is a banded Toeplitz matrix. From Lemma 1--3 of \cite{zhang2022faster}, $\bm{G}$ is positive definite but has close-to-zero eigenvalues. Moreover, the matched-filter output noise, $\eta(t)$,  is also correlated. Its correlation properties are given by
\begin{align}
    \mathbb{E}[\eta_i\eta_j^*]
    &=N_0 g\!\left((i-j)\delta T\right)=N_0g_{i-j}, \\
    \mathbb{E}[\bm\eta\bm\eta^{\dagger}]
    &=N_0\bm G.
\end{align}

Finally, the transmission power level is assumed to be $P$, and we have the relationship 
    $P=\frac{E_s}{\delta T}$.

\section{Likelihood Formulation and Full-State AIR}
To investigate the AIR of FTN signaling with finite constellations, we first need to establish the likelihood and symbol posteriors used in its evaluation. The full-state BCJR provides exact symbol posteriors. When $\bm G$ is positive definite, the likelihood resulting from \eqref{eqn:sysmodelunger} is proportional to
% For decoding a sequence of FTN symbols, ML detection is the optimal choice. 
% The ML objective is to find the candidate sequence
% $\bm{a}$ that maximizes the likelihood $p(\bm{y}|\bm{a})$. When $\bm{G}$ is
% positive definite, the likelihood resulting from \eqref{eqn:sysmodelunger} is
% proportional to
\begin{align}
    p(\bm{y}|\bm{a})
    \propto
    \exp\left(
    -\frac{1}{N_0}
    (\bm{y}-\bm{G}\bm{a})^\dagger
    \bm{G}^{-1}
    (\bm{y}-\bm{G}\bm{a})
    \right).
    \label{eqn:likelihoodabovthres}
\end{align}
An exhaustive search would require evaluating all $M^N$ candidate sequences and
is therefore computationally prohibitive. Instead, trellis-based sequence
estimation evaluates the likelihood recursively.
The input symbols are independent, and the equiprobable input
distribution used in the AIR analysis assigns the same prior
probability to every candidate sequence. Under this distribution,
sequence ML and sequence MAP detection are equivalent. The BCJR
algorithm addresses the corresponding symbol-wise marginal problem.
Using the sum-product recursion, it evaluates the posterior
probability of each symbol value and makes the decision by 
\begin{align}
    \hat a_k
    =
    \underset{a_k\in\mathcal{A}}{\arg\max}
    \;p(a_k|\bm{y}),
    \qquad k=0,\ldots,N-1.
    \label{eqn:maprule}
\end{align}
With the full state, this marginalization is exact. When the state is
reduced, some symbol histories are merged and the resulting soft
metric becomes an approximation to the full-state posterior. Because
the finite-memory metric depends on $L$ past interfering symbols, a
full-state trellis contains $M^L$ states. A receiver that retains only
$J<L$ of these symbols and estimates the remaining interference through
traceback or decision feedback is referred to as a reduced-state
receiver.

We use the AIR to evaluate the
information rate delivered by the receiver's soft-output metric for a
specified constellation, SNR, and state length. The AIR quantifies the maximum information rate achievable by the receiver, assuming the use of an optimal channel code, without requiring the design or simulation of a specific coding scheme. It is typically expressed in bits per transmitted symbol.

In FTN signaling, for sufficiently aggressive acceleration, matrix $\bm G$ becomes rank-deficient \cite{zhang2025iapr}.
If the matrix $\bm{G}$ becomes rank-deficient, the inverse in
\eqref{eqn:likelihoodabovthres} no longer exists. 
Since $\bm{G}$ is the
correlation matrix composed of the matched filter samples, rank deficiency means that
the samples contain fewer statistically independent dimensions. In addition,
because the noise covariance matrix is $N_0\bm{G}$, the noise has the same
reduced-dimensional support.
This shows that under sufficient acceleration, in FTN detection problem, the noise is singular, and the noiseless samples are also singular. 
Therefore, in order to show that the FTN detection problem is well-defined, and to validate that FTN can still use ML detection, the problem
must be analyzed on the range of $\bm{G}$ rather than in the full
$N$-dimensional space. On the other hand, since the likelihood of FTN detection should be formulated in a different way, the application of the BCJR decoder to FTN also requires validation. Consequently, neither the likelihood in \eqref{eqn:likelihoodabovthres} nor
its Ungerboeck factorization can be assumed to hold, and the validity of the
BCJR receiver for FTN must be proved rather than taken for granted.

In this section, we first bring in RKHS as a tool to facilitate the analysis, 
we then show that the noiseless
received signal, the noise, and the received vector all lie in the range of
$\bm G$. 
We introduce the full-state BCJR detector and formulate the likelihood using Moore--Penrose inverse on this lower-dimensional sub-space for low-rank $\bm{G}$. We eventually 
show that the resulting likelihood produces the same Ungerboeck branch metric, which means the BCJR detector can be directly applied to FTN with high acceleration rate.

\subsection{A Primer on RKHS}\label{subsec:rkhssubsec}

% To rigorously describe this situation, we use the RKHS associated with the noise covariance matrix $N_0\bm{G}$. Since
% $N_0$ is only a positive scaling factor, it does not change the underlying
% space, we therefore omit this factor in the RKHS definition. The RKHS is defined as
% follows.
To formulate detection with singular noise, we work within
the subspace supporting the noise distribution. The RKHS
associated with the noise covariance defines a covariance-weighted
geometry on this subspace through its inner product and norm.
We bring in these concepts to prove the validity of the
Ungerboeck metric for FTN reception.
Since the positive scaling factor $N_0$ does not change the
underlying space, we define the RKHS using $\bm{G}$ and
retain $N_0$ explicitly in the likelihood. The RKHS is
defined as follows.

\textit{Definition \cite{rkhspaper}:} An RKHS $\mathcal{H}({\bm{G}})$ associated with
a covariance matrix ${\bm{G}}$  is a Hilbert space  with inner product $\langle\cdot,\cdot\rangle_{\mathcal{H}}$
satisfying the reproducing property
\begin{align}
    \left\langle {\bm{G}}[:,k],\bm m\right\rangle_{\mathcal{H}}
    =m_k, \quad k\in\{0,1,\dots,N-1\},
    \label{eqn:reproducingkernel}
\end{align}
for every $\bm m\in\mathcal{H}({\bm{G}})$. The vector
${\bm{G}}[:,k]$ is called the reproducing kernel, and it satisfies
\begin{align}
    \bm{G}[:,k]\in\mathcal{H}({\bm{G}}). 
\end{align}

From now on, we use the notations $\mathcal{H}$ and
$\mathcal{H}({\bm{G}})$ interchangeably. 
In the finite-dimensional FTN model, 
the range of $\bm G$ represents the set of observation vectors supported by the
FTN signal and the noise model.

For vectors $\bm{b},\bm{c}\in\mathcal{H}$, when $\bm{G}$ is nonsingular, the
RKHS inner product is defined as \cite{rkhspaper}
\begin{align}
    \left\langle\bm{b},\bm{c}\right\rangle_{\mathcal{H}}
    =\bm{b}^{\dagger}{\bm{G}}^{-1}\bm{c}.
    \label{eqn:innprodrkhs}
\end{align}
When the noise covariance matrix $\bm{G}$ is rank-deficient, the inner product
is instead defined using the pseudo inverse or the Moore--Penrose inverse of $\bm G$, which is denoted as ${\bm{G}}^{\#}$, and  is given as  \begin{align}
    \left\langle\bm{b},\bm{c}\right\rangle_{\mathcal{H}}
    =\bm{b}^{\dagger}{\bm{G}}^{\#}\bm{c}.
    \label{eqn:innprodrkhssingular}
\end{align}  The matrix ${\bm{G}}^{\#}$ acts as the
inverse on the nonzero-eigenvalue subspace and discards the zero-variance
directions.
 The RKHS norm is defined as
\begin{align}
    \lVert\bm{b}\rVert^2_{\mathcal{H}}
    \triangleq
    \left\langle\bm{b},\bm{b}\right\rangle_{\mathcal{H}}.
\end{align}
For a singular matrix, this definition is an inner product on the range of
$\bm{G}$. 
Since the reproducing kernels of
$\mathcal{H}$ are the column vectors of $\bm{G}$, we then have the following proposition.
\begin{proposition}\label{prop:rkhscolumnspace}
    If $\bm{b}\in\mathcal{H}$, then
    $\bm{b}\in\mathcal{R}({\bm{G}})$. Equivalently, $\bm{b}$ belongs to the space
    spanned by the columns of ${\bm{G}}$. Conversely, every vector in
    $\mathcal{R}({\bm{G}})$ belongs to $\mathcal{H}$.
\end{proposition}

\begin{proof}
    The proof can be found in \cite{proofforpropo1}. \hfill$\square$
\end{proof}

\subsection{Application to FTN Detection}
We now proceed to proving the fact that the FTN detection problem under severe correlation is well-defined.
\begin{proposition}\label{prop:nonsingulardetect}
    In FTN transmission, even when the noise covariance matrix
    $N_0\bm{G}$ is singular, the detection problem is not singular.
\end{proposition}

\begin{proof}
    From \eqref{eqn:sysmodelunger}, the noiseless received vector
    $\bm{G}\bar{\bm{a}}$ is a linear combination of the columns of $\bm{G}$.
    Therefore,
    \begin{align}
        \bm{G}\bar{\bm{a}}\in\mathcal{R}({\bm{G}}).
    \end{align}
    Moreover, since the noise covariance matrix is $N_0\bm{G}$, the noise has
    zero variance in every eigendirection corresponding to a zero eigenvalue
    of $\bm{G}$. Consequently,
    \begin{align}
        \bm{\eta}\in\mathcal{R}({\bm{G}}),
        \qquad\text{almost surely}. \quad\quad\quad\quad\hfill \square
    \end{align}
This proposition is essential for establishing the theoretical foundations for FTN detection under aggressive acceleration. Trellis-based decoders evaluated in this paper, namely, the Ungerboeck-type decoder and BCJR receiver, cannot
be transferred to the FTN setting by simply replacing its signal model,
because the FTN correlation matrix may be rank deficient and the
full-dimensional Gaussian likelihood is then undefined. The receiver
must instead be derived on $\mathcal{R}(\mathbf G)$. 
    
    By Proposition~\ref{prop:rkhscolumnspace}, the correlated transmitted samples and the noise both belong to the same RKHS almost surely, their linear combination also resides in the same space \cite{rkhspaper}.
    It follows that
    \begin{align}
        \bm{y}\in\mathcal{R}({\bm{G}}),
        \qquad\text{almost surely}.
        \notag
    \end{align}
     This common
    support is a consequence of the FTN matched filter model, in which the
 noise is generated through the matrix $N_0\bm{G}$ and the symbols $\bar{\bm{a}}$ are projected onto the space spanned by columns of $\bm{G}$ due to correlation.
    The key finding of this section is that a singular FTN noise covariance
matrix does not invalidate the detection problem. Although $\bm G$ may be
rank-deficient, the noiseless signal and the noise share the same support,
$\mathcal{R}(\bm G)$. Consequently, a valid likelihood can still be defined.
% this subspace. This result provides the theoretical foundation for applying
% the Ungerboeck-type receiver to strongly accelerated FTN signaling.
% Therefore, the singularity of the covariance matrix does not make the
%     detection problem ill-defined. 
This common support permits a likelihood formulation on $\mathcal{R}(\bm G)$, providing the basis for evaluating the symbol posteriors and AIR even when the matched-filter noise covariance is singular.
    If a signal component existed outside
    $\mathcal{R}({\bm{G}})$, this common-support argument would no longer
    apply.
\end{proof}

\subsection{Symbol Posterior Evaluation }
Since the FTN detection problem is a valid detection problem, we now introduce the full-state BCJR detector structure.  
The BCJR decoder makes symbol decisions according to the MAP rule in \eqref{eqn:maprule}.
 The trellis state
$\sigma_k$ is defined as
\begin{align}
    \sigma_k
    =
    (a_{k-1},a_{k-2},\ldots,a_{k-L}),
    \qquad k=0,\ldots,N.
\end{align}
Thus, $\sigma_k$ contains only the past interfering symbols. For transmission without a cyclic prefix, $\sigma_0$ is a fixed
boundary state. Equivalently, the symbols preceding the block can be set to
zero, such that
    $\sigma_0=(a_{-1},a_{-2},\ldots,a_{-L})=(0,\ldots,0)$.
The $N$ transmitted symbols produce $N+1$ trellis state positions,
    $\bm{\sigma}=    [\sigma_0,\sigma_1,\ldots,\sigma_N]^T$. 
Accordingly, $\sigma_k$ and $a_k$ together determine $\sigma_{k+1}$. 
The symbol posterior can be written as
\begin{align}
    p(a_k|\bm{y})
    =
    \sum_{\bm{\sigma}}
    \sum_{\bm{a}\backslash a_k}
    p(\bm{a},\bm{\sigma}|\bm{y}), \label{eqn:insert12}
\end{align}
where $\bm{a}\backslash a_k$ denotes all candidate symbols except $a_k$. Furthermore,
\begin{align}
    p(\bm{a},\bm{\sigma}|\bm{y})
    &\propto
    p(\bm{a},\bm{\sigma},\bm{y}) \\
    &\propto
    p(\bm{y}|\bm{a},\bm{\sigma})
    p(\bm{\sigma}|\bm{a})
    p(\bm{a}). \label{eqn:insert14}
\end{align}
% Because the trellis transition is determined by $\sigma_k$, $a_k$, and
% $\sigma_{k+1}$,
% \begin{align}
%     p(\bm{\sigma}|\bm{a})
%     =
%     \prod_{k=0}^{N-1}
%     p(\sigma_{k+1}|\sigma_k,a_k).
%     \label{eqn:assemble1}
% \end{align}
% We denote the transition probability $p(\sigma_{k+1}|\sigma_k,a_k)$ by
% $T(\sigma_k,a_k,\sigma_{k+1})$. It equals one when the branch
% $\sigma_k\xrightarrow{a_k}\sigma_{k+1}$ exists and equals zero otherwise.
Given the current state $\sigma_k$ and input symbol $a_k$,
the next state $\sigma_{k+1}$ is independent of all earlier
states and inputs, we have 
\begin{align}
p(\bm{\sigma}\mid\bm{a})
= \prod_{k=0}^{N-1} p(\sigma_{k+1}\mid\sigma_k,a_k) 
= \prod_{k=0}^{N-1} T(\sigma_k,a_k,\sigma_{k+1}).\label{eqn:assemble1}
\end{align}
Here, $T(\sigma_k,a_k,\sigma_{k+1})$ is the transition
indicator, which equals one, if input $a_k$ produces a valid
transition from $\sigma_k$ to $\sigma_{k+1}$, and zero otherwise.
Moreover,
\begin{align}
    p(\bm{y}|\bm{a},\bm{\sigma})
    =
    p(\bm{y}|\bm{a}), \label{eqn:likelihoodindep}
\end{align}
since once the candidate symbols $\bm a$ are determined, the state sequence $\bm \sigma$ is also determined, the conditional likelihood $p(\bm{y}|\bm{a},\bm{\sigma})$ is therefore irrelevant to $\bm \sigma$. 
Following the Ungerboeck observation model in
\cite{ungerboeckfirstpaper}, the likelihood in \eqref{eqn:likelihoodindep} should be decomposed into branch
metrics to enable trellis-based iterative computation for the likelihood
\begin{align}
    p(\bm{y}|\bm{a})
    \propto
    \prod_{k=0}^{N-1}
    \mathcal{J}(a_k,\sigma_k),   \label{eqn:likelihooddecompfullstate}
\end{align}
where the full-state branch metric is
\begin{align}
&\mathcal{J}(a_k,\sigma_k)
=
\exp\bigg\{
\frac{1}{N_0}\Re\bigg[
2y_ka_k^*
-|a_k|^2g_0
\notag\\[-0.3ex]
&\qquad
-2\sum_{l=1}^{L}a_k^*a_{k-l}g_l
\bigg]\bigg\},
\qquad k=0,\ldots,N-1. \label{eqn:branchmetricfulls}
\end{align}
For $k-l<0$, the corresponding symbols are determined by the boundary
condition. This branch metric is valid when the noise covariance matrix is invertible, or when \eqref{eqn:likelihoodabovthres} exists. 
However, as analyzed in 
\cite{singularmultivardistbook}, when the noise covariance matrix is singular,
the likelihood function is proportional to
\begin{align}
    p(\bm{y}\mid\bm{a})
    &\propto
    \exp\left(
    -\frac{1}{N_0}
    (\bm{y}-\bm{G}\bm{a})^\dagger
    \bm{G}^{\#}
    (\bm{y}-\bm{G}\bm{a})
    \right).
    \label{eqn:likelihoodbelowthres}
\end{align}
Note that,
the correlated noise and every noiseless candidate  
vector $\bm G\bm a$ lie in the range of $\bm G$, while they both have no
component in the orthogonal null space. The Moore--Penrose inverse is
therefore used to define the RKHS geometry on the supported
subspace. The singularity changes the geometry in which the likelihood
is evaluated, but it does not invalidate the branch metric in \eqref{eqn:branchmetricfulls}.
The following theorem establishes the likelihood factorization used in the AIR analysis.
% Equation~\eqref{eqn:likelihoodbelowthres} leads to the first and principal
% theorem of this paper.

\begin{theorem}\label{thm:metricworks}
    Even when $\bm{G}$ is singular, the Ungerboeck branch metric used to
    maximize the likelihood remains valid. When $a_k$ labels the transition
    $\sigma_k\rightarrow\sigma_{k+1}$, the branch metric is
    \begin{align}
        J(a_k,\sigma_k)
        =
        2\text{Re}\{a^*_ky_k\}
        -|a_k|^2g[0]
        -2\text{Re}\left\{
        a_k^*\sum_{l=1}^{L}g[l]a_{k-l}
        \right\}. \label{eqn:metricstillvalid}
    \end{align}
\end{theorem}

\begin{proof}
    See Appendix~\ref{app:metricworks} for the proof.
\end{proof}
% Therefore, we have shown that under any acceleration rate, the branch metric for computing the likelihood is always valid, this enables the direct application of the BCJR decoder to FTN detection. 

\begin{figure}
    \centering
    \includegraphics[width=0.9\linewidth]{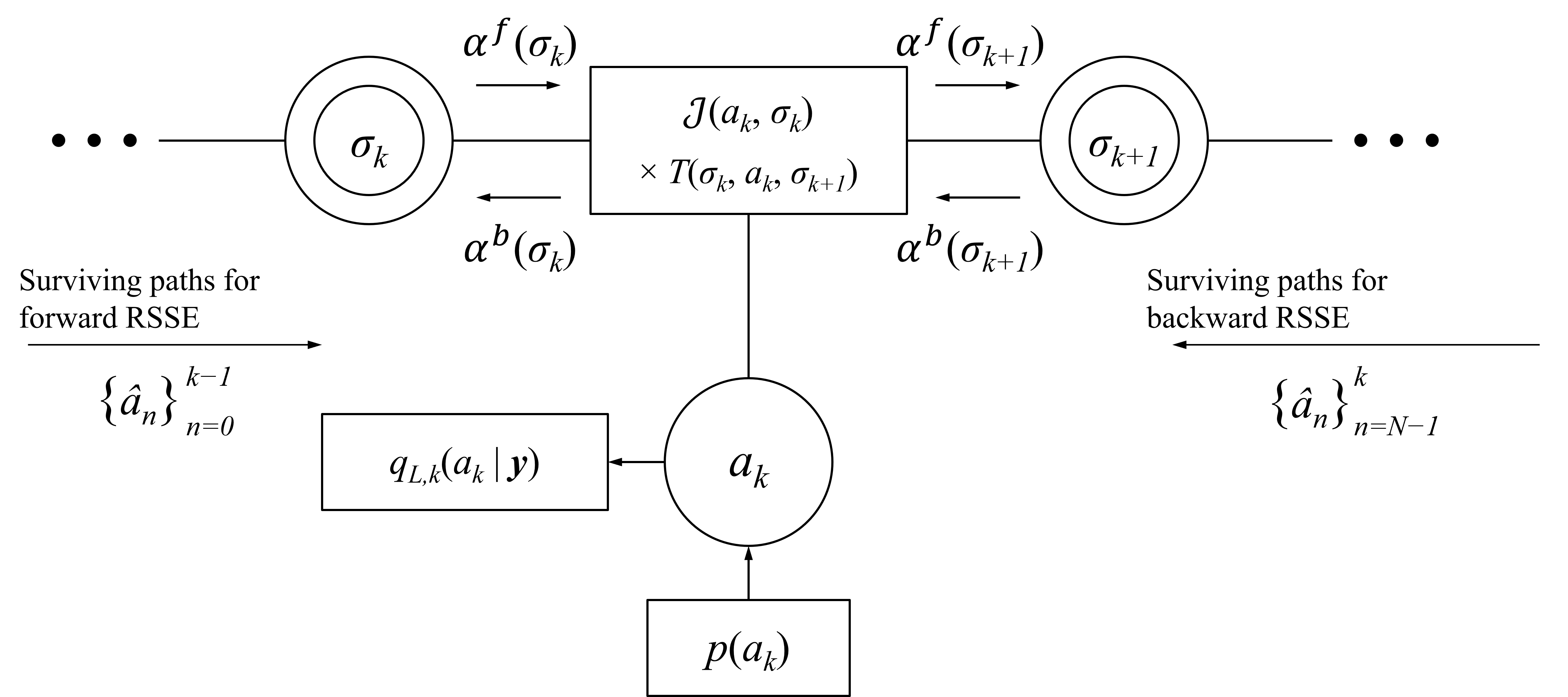}
    \caption{Factor graph and message passing scheme for the FTN BCJR detector.}
    \label{fig:factorgraph}
\end{figure}

This result allows the symbol posteriors required for AIR evaluation to be computed using the Ungerboeck factorization even when the noise covariance is singular.
In the full-state trellis, all $L$ past symbols are
contained in $\sigma_k$.
Since the symbols are i.i.d.,
\begin{align}
    p(\bm{a})
    =
    \prod_{k=0}^{N-1}p(a_k).
    \label{eqn:assemble3}
\end{align}
Inserting \eqref{eqn:assemble1}, \eqref{eqn:likelihooddecompfullstate}, and \eqref{eqn:assemble3} into \eqref{eqn:insert14}, and \eqref{eqn:insert14} into \eqref{eqn:insert12} results in
\begin{align}
    p(a_k|\bm{y})
    \propto
    \sum_{\bm{\sigma}}
    \sum_{\bm{a}\backslash a_k}
    \prod_{i=0}^{N-1}
    \mathcal{J}(a_i,\sigma_i)
    T(\sigma_i,a_i,\sigma_{i+1})
    p(a_i).
    \label{eqn:decomposebeforesumproduct}
\end{align}
In order to simplify the computation, \eqref{eqn:decomposebeforesumproduct}
can be evaluated recursively using the sum-product algorithm 
\cite{sumproductalgorithm}. The factor graph for the message-passing
implementation is shown in Fig.~\ref{fig:factorgraph}.

When the operator $\sum_{\sim \{\sigma_{k+1}\}}$ denotes summation over all variables except $\sigma_{k+1}$, the forward message is initialized at the boundary state $\sigma_0$ and is
defined by
\begin{align}
    \alpha^f(\sigma_{k+1})=\sum_{\sim\{\sigma_{k+1}\}}\alpha^f(\sigma_{k})\mathcal{J}(a_k,\sigma_{k})T(\sigma_k,a_k,\sigma_{k+1})p(a_k). \label{eqn:forwardmsgupdate}
\end{align}
The backward message is initialized at the terminal state $\sigma_N$ and is
defined by
\begin{align}
    \alpha^b(\sigma_k)=\sum_{\sim\{\sigma_k\}}\alpha^b(\sigma_{k+1})\mathcal J(a_k,\sigma_k)T(\sigma_k,a_k,\sigma_{k+1})p(a_k). \label{eqn:backwardmsgupdate}
\end{align}
The symbol posterior is then obtained as
\begin{align}
    p(a_k|\bm{y})\propto q_{L,k}(a_k;\bm y)= p(a_k)\sum_{\sim\{a_k\}}\alpha^f(\sigma_{k})\alpha^b(\sigma_{k+1})\notag \\
    \times\mathcal J(a_k,\sigma_k)T(\sigma_k,a_k,\sigma_{k+1}), \label{eqn:posteriorandscorefull}
\end{align}
where we use $q_{L,k}(a_k;\bm y)$ to denote the output score for $a_k$ of the full-state BCJR receiver of state length $L$.
The receiver chooses the symbol with the largest posterior as in \eqref{eqn:maprule}.

\subsection{Full-State AIR and Its High-SNR Limit}
For independent, equiprobable input symbols, the AIR expression
evaluated in this section is \cite[(54)]{Gokhanmismatch}
\begin{align}
 \text{AIR}_{L}(\delta,N_0)
 &=\log_2 M+\frac{1}{N}\sum_{k=0}^{N-1}
 \mathbb E\!\left[
 \log_2
 \frac{q_{L,k}(\bar a_k;\bm y)}
      {\sum_{c\in\mathcal A}q_{L,k}(c;\bm y)}
 \right].
 \label{eq:ftn_air_definition}
\end{align}
The expectation is taken over the transmitted block and the
correlated noise. In numerical evaluation, the receiver computes
the scores for all candidate symbols, and the numerator selects
the score assigned to the actual transmitted symbol $\bar a_k$.
This symbol need not coincide with the receiver decision
$\hat a_k$. 
% We retain the signed expression in
% \eqref{eq:ftn_air_definition} for the subsequent analysis;
% its nonnegative part gives the corresponding achievable rate.

For the full-state receiver implemented using the exact
sum-product algorithm, all $L$ past interfering symbols are
included in the state. Under the adopted finite-memory model,
the likelihood factorization in
\eqref{eqn:likelihooddecompfullstate} and the recursion in
\eqref{eqn:posteriorandscorefull} perform exact
marginalization over the candidate sequences. To make this
connection explicit, define
\[
 \mathcal A_k(c)
 \triangleq
 \{\bm a\in\mathcal A^N:a_k=c\}.
\]
Using the full-state sum-product expression in
\eqref{eqn:posteriorandscorefull}, 
%with the likelihood in \eqref{eqn:likelihoodbelowthres}, 
the full-state score can
be written as
\begin{align}
 q_{L,k}(c;\bm y)
 &\propto
 \sum_{\bm a\in\mathcal A_k(c)}p(\bm y\mid\bm a).
 % \notag\\
 % &\propto
 % \sum_{\bm a\in\mathcal A_k(c)}
 % \exp\!\left(
 % -\frac{\|\bm y-\bm G\bm a\|_{\mathcal H}^{2}}{N_0}
 % \right).
 \label{eq:ftn_air_exact_metric}
\end{align}
% The second expression follows from the RKHS likelihood
% representation. 
Each sum fixes $a_k=c$ and ranges over all
$M^{N-1}$ choices of the remaining symbols. The proportionality
factors may depend on $k$ and $\bm y$, but not on $c$.
Since $p(\bm a)=M^{-N}$, Bayes' rule gives
\begin{align}
 p(\bar a_k=c\mid\bm y)
 &=\sum_{\bm a\in\mathcal A_k(c)}p(\bm a\mid\bm y)
 \notag\\
 &=\frac{
 \sum_{\bm a\in\mathcal A_k(c)}p(\bm y\mid\bm a)p(\bm a)
 }{
 \sum_{\bm a\in\mathcal A^N}p(\bm y\mid\bm a)p(\bm a)
 }
 \notag\\
 &=\frac{
 \sum_{\bm a\in\mathcal A_k(c)}p(\bm y\mid\bm a)
 }{
 \sum_{\bm a\in\mathcal A^N}p(\bm y\mid\bm a)
 }.
 \label{eq:ftn_air_exact_posterior}
\end{align}
% Moreover, the sets $\mathcal A_k(c)$, $c\in\mathcal A$,
% form a disjoint partition of $\mathcal A^N$, so
% \[
%  \sum_{d\in\mathcal A}
%  \sum_{\bm a\in\mathcal A_k(d)}p(\bm y\mid\bm a)
%  =\sum_{\bm a\in\mathcal A^N}p(\bm y\mid\bm a).
% \]
% Consequently, normalizing the scores in
% \eqref{eq:ftn_air_exact_metric} cancels their common
% proportionality factor and yields
% \begin{equation}
%  p_{L,k}(c\mid\bm y)
%  =\frac{q_{L,k}(c;\bm y)}
%         {\sum_{d\in\mathcal A}q_{L,k}(d;\bm y)}
%  =p(\bar a_k=c\mid\bm y).
%  \label{eq:ftn_air_matched_posterior}
% \end{equation}
% Thus, the normalized full-state sum-product output is the
% exact symbol posterior.
Since the sets $\mathcal A_k(d)$, $d\in\mathcal A$, form a
disjoint partition of $\mathcal A^N$,
\begin{equation}
 \sum_{\bm a\in\mathcal A^N}p(\bm y\mid\bm a)
 =
 \sum_{d\in\mathcal A}
 \sum_{\bm a\in\mathcal A_k(d)}p(\bm y\mid\bm a).
 \label{eq:ftn_air_partition}
\end{equation}
Furthermore, \eqref{eq:ftn_air_exact_metric} implies that,
for fixed $k$, there exists a factor $\kappa(\bm y)>0$,
independent of $c$, such that
\begin{equation}
 q_{L,k}(c;\bm y)
 =
 \kappa(\bm y)
 \sum_{\bm a\in\mathcal A_k(c)}p(\bm y\mid\bm a).
 \label{eq:ftn_air_score_factor}
\end{equation}

Substituting \eqref{eq:ftn_air_partition} and
\eqref{eq:ftn_air_score_factor} into the final fraction in
\eqref{eq:ftn_air_exact_posterior} gives
\begin{align}
 p(\bar a_k=c\mid\bm y)
 &=
 \frac{\kappa^{-1}(\bm y)\,q_{L,k}(c;\bm y)}
      {\kappa^{-1}(\bm y)
       \sum_{d\in\mathcal A}q_{L,k}(d;\bm y)}
 \notag\\
 &=
 \frac{q_{L,k}(c;\bm y)}
      {\sum_{d\in\mathcal A}q_{L,k}(d;\bm y)}
 % \notag\\
 % &=p_{L,k}(c\mid\bm y),
 \label{eq:ftn_air_matched_posterior}
\end{align}
% where the last equality follows from
% \eqref{eq:ftn_air_normalized_metric}.
Thus, the full-state sum-product receiver produces the
exact symbol posterior.
Finally, substituting \eqref{eq:ftn_air_matched_posterior} into
\eqref{eq:ftn_air_definition} we have
\begin{align}
 \text{AIR}_{L}(\delta,N_0)
 &=\log_2 M+\frac{1}{N}\sum_{k=0}^{N-1}
 \mathbb E\!\left[\log_2 p(\bar a_k\mid\bm y)\right]
 \notag\\
 &=\log_2 M-\frac{1}{N}\sum_{k=0}^{N-1}
 H(\bar a_k\mid\bm y)
 \notag\\
 &=\frac{1}{N}\sum_{k=0}^{N-1}I(\bar a_k;\bm y),
 \label{eq:full_state_entropy}
\end{align}
which shows that the full-state AIR is equal to the average mutual information
between $\bar a_k$ and $\bm{y}$. Since the entropy of the input symbols is $H(\bar a_k)=\log_2 M$,
\begin{equation}
 0\leq\text{AIR}_{L}(\delta,N_0)\leq\log_2 M.
 \label{eq:full_state_bounds}
\end{equation}
% As SNR increases, the full-state BCJR decoder can better distinguish between sequences since ML or MAP detection is optimum, and full-state BCJR directly reflects the likelihood of the sequence. Eventually, as SNR approaches infinity, BCJR decoder will be able to correctly decode any sequence.  We then have the following theorem
Equation \eqref{eq:full_state_entropy} relates the full-state AIR to the remaining uncertainty about each transmitted symbol. When distinct transmitted sequences have a positive minimum RKHS distance, this uncertainty vanishes as the noise power decreases, giving the following high-SNR limit.
\begin{theorem}\label{thm:airfulllogm}
    For the full-state BCJR decoder, as SNR approaches infinity, we have 
    \begin{align}
        \lim_{N_0\rightarrow0}
\text{AIR}_{L}(\delta,N_0)
=
\log_2 M .
    \end{align}
\end{theorem} 
\begin{proof}
    Assume that the finite constellation satisfies
\begin{equation}
d_{\min}^2(\delta)
\triangleq
\min_{\bm a\neq\bm b}
\left\|\bm G(\bm a-\bm b)\right\|_{\mathcal H}^{2}
>0.
\label{eq:full_state_dmin}
\end{equation}
After whitening the observation on the range of $\bm G$,
the distance between any two noiseless sequences is
$\left\|\bm G(\bm a-\bm b)\right\|_{\mathcal H}$.
Therefore, the sequence error probability of the maximum-likelihood
receiver satisfies
\begin{equation}
P_{\mathrm e}
\leq
(M^N-1)
Q\!\left(
\frac{d_{\min}(\delta)}
{\sqrt{2N_0}}
\right)
\longrightarrow 0,
\qquad N_0\rightarrow0 .
\label{eq:full_state_error}
\end{equation}
By Fano's inequality,
\begin{equation}
H(\bar{\bm a}\mid\bm y)
\leq
h_2(P_{\mathrm e})
+
P_{\mathrm e}\log_2(M^N-1)
\longrightarrow0 .
\label{eq:full_state_fano}
\end{equation}
Since
\[
H(\bar a_k\mid\bm y)
\leq
H(\bar{\bm a}\mid\bm y),
\]
substituting into \eqref{eq:full_state_entropy} gives
\begin{equation}
\lim_{N_0\rightarrow0}
\text{AIR}_{L}(\delta,N_0)
=
\log_2 M .   \quad\quad\quad\quad\hfill \square
\label{eq:full_state_limit}
\end{equation}
\end{proof}

% For $J<L$, the receiver replaces the exact treatment of the
% discarded symbol histories with survivor-path traceback.
% In particular, the terms with $m=J+1,\ldots,L$ use the
% estimates $\dot a_{k-m}$, while bias correction relies on
% tentative future-symbol decisions. When a max-log implementation
% is used, summations are additionally approximated by
% maximizations. The resulting normalized scores
% $p_{J,k}(c\mid\bm y)$ therefore need not equal the true
% symbol posteriors. Accordingly,
% \eqref{eq:ftn_air_definition}, evaluated using the actual
% receiver outputs, measures the AIR associated with this
% mismatched metric. The following analysis relates the loss
% in AIR to the scores assigned to the transmitted symbol and
% its competing hypotheses.

\section{Reduced-State Receiver for AIR Evaluation}
\label{sec:decodersecAfullstate}

% For the reduced-state BCJR decoder, we adapt the reduced-state Ungerboeck-type receiver of \cite{Gokhanhocampaper} to FTN signaling. This receiver is selected because its state length and feedback iterations provide adjustable complexity, its observation model avoids explicit noise whitening, and its soft symbol outputs support mismatched-decoding AIR evaluation. The adaptation expresses the branch metrics and feedback terms through the FTN pulse-correlation coefficients, thereby allowing the effects of time packing and receiver complexity to be investigated within the same framework.

% This section presents the Ungerboeck-type BCJR receiver for FTN signaling. We describe the reduced-state implementation, including survivor-path traceback and bias correction using bidirectional decision feedback.

To examine the AIR attainable with limited receiver complexity, we use the reduced-state Ungerboeck-type receiver of \cite{Gokhanhocampaper}, adapted to the FTN scheme. Its state length and feedback iterations provide adjustable complexity, which can be made as close to ML detection as desired, while its final symbol scores are used to evaluate the AIR. In this section, we specify the traceback, bidirectional feedback, and max-log recursions to examine how the resulting approximations affect the achievable rate.

\subsection{Reduced-State BCJR Decoder}\label{sec:decodersecBredstate}

In reduced-state detection, only $J$ of the $L$ interfering symbols are involved
in the trellis state, where $0\leq J\leq L$. At time $k$, the reduced state is
    $\sigma_k=(a_{k-1},\ldots,a_{k-J})$. 
The symbols outside the reduced state are supplied by traceback. At step $k$, the interfering symbols prior to $k-J$ are chosen by tracing back the optimal path leading to $\sigma_{k-1}$. Specifically,
the traceback provides the estimates
    $\dot a_{k-J-1},\ldots,\dot a_{k-L}$.
Similar to \eqref{eqn:branchmetricfulls}, the reduced-state branch metric is 
\begin{align}
&\tilde{\mathcal{J}}(a_k,\sigma_k)
=
\exp\bigg\{
\frac{1}{N_0}\Re\bigg[
y_ka_k^*
-\frac{1}{2}|a_k|^2g_0
\notag\\[-0.3ex]
&\qquad
-\sum_{l=1}^{J}a_k^*a_{k-l}g_l
-\sum_{m=J+1}^{L}a_k^*\dot a_{k-m}g_m
\bigg]\bigg\}. \label{eqn:reducestatemetric}
\end{align}
The first summation uses the symbols in the reduced state, whereas
the second summation uses the traceback estimates for the discarded past
symbols. Thus, $\widetilde{\mathcal{J}}(a_k,\sigma_k)$
is the branch-metric contribution associated with the candidate symbol
$a_k$ and the state $\sigma_k$. It quantifies how well this candidate
transition explains the received sample $y_k$ under the Ungerboeck
observation model.

Although traceback approximates the full-state history, the future symbols are another source of interference. In reduced-state Ungerboeck-type
receivers, this interference is referred to as the bias
\cite{starkpaper,Gokhanhocampaper}. The bias results from the correlation
between the traceback past symbols and the future symbols that are not
included in the current reduced state. The bias is estimated using future
symbol decisions obtained from the backward process
\begin{align}
B_k&\left(
\{\dot a_i\}_{i=k-L+1}^{k-J},
\{\hat a_i\}_{i=k+1}^{k+L-J}
\right)
\notag\\
&=
2\sum_{m=k-L+1}^{k-J}
\mathrm{Re}
\left\{
\dot a_m^{*}
\sum_{l=k-m+1}^{L}
g_l\hat a_{m+l}
\right\}.
\label{eqn:bias_definition}
\end{align}
At the block boundaries, only valid symbol indices are included in the 
corresponding sums. 
The symbols $\{\dot a_i\}_{i=k-L+1}^{k-J}$ are the traceback
estimates of the discarded past symbols, whereas
$\{\hat a_i\}_{i=k+1}^{k+L-J}$ are tentative future-symbol decisions
obtained from the backward process. For each discarded past symbol
$\dot a_m$, the inner summation evaluates its correlation with the
future symbols $\hat a_{m+l}$ that occur after time $k$. Consequently,
$B_k\left(
\{\dot a_i\}_{i=k-L+1}^{k-J},
\{\hat a_i\}_{i=k+1}^{k+L-J}
\right)$ measures the anticausal interference produced by the
past-future symbol interactions that are not included in the
reduced-state branch metric. It is a correction to the metric rather
than an additional noise term.
The bias term is not present in the full-state detection problem since in full-state detection, the symbols $\{\dot a_i\}_{i=k-L+1}^{k-J}$ are included in the states,  its correlation with the future symbols $\{\hat a_i\}_{i=k+1}^{k+L-J}$ is accounted for in the trellis updating process. 

For each candidate transition from $\sigma_k$ to $\sigma_{k+1}$, the survivor
path is selected by maximizing the bias-corrected branch metric
\begin{align}
    \hat\sigma_{k-1}=&\underset{\sigma_{k-1}}{\arg\max}~ \ln \tilde{\mathcal J}(a_k,\sigma_{k-1}) \notag\\
    &\quad-B_k\left(
\{\dot a_i\}_{i=k-L+1}^{k-J},\{\hat a_i\}_{i=k+1}^{k+L-J}
\right)/N_0. \label{eqn:decisionbias}
\end{align}
The bias is used for the survivor-path selection. It is not included in the
forward or backward message updates.

Here we use the logarithm on the metric to avoid numerical explosion brought by production due to long sequences. 
Let $\Lambda_k$ denote the
accumulated log-metric of the survivor path that reaches state
$\sigma_k$. We initialize
$\Lambda_0=0$ and define
\begin{equation}
    \Lambda_k
    =
    \sum_{i=0}^{k-1}
    \ln \tilde{\mathcal{J}}(a_i,\sigma_i),
    \qquad k\geq 1.
    \label{eqn:redstatelikelihood}
\end{equation}
 The bias-corrected metric
in \eqref{eqn:decisionbias} is used to select the survivor $\hat{\sigma}_k$. After
this survivor has been selected, appending the candidate symbol $a_k$
adds the log-metric contribution of the new branch
\begin{equation}
    \Lambda_{k+1}
    =
    \Lambda_k
    +
    \ln \tilde{\mathcal{J}}(a_k,\hat{\sigma}_k).
    \label{eqn:accumulated_metric_update}
\end{equation}
The log-domain sum-product recursions are
approximated by maximization. For valid trellis transitions, for forward and backward passes, we have
\begin{align}
&\ln\alpha^f(\sigma_{k+1})
=
\max_{\sigma_k}
\bigg[
\ln\alpha^f_k(\sigma_k)
+\ln\tilde{\mathcal{J}}(a_k,\sigma_k)
+\ln p(a_k)
\bigg], \label{eqn:logforwardmsgupdate}
\\
&\ln\alpha^b(\sigma_k)
=
\max_{\sigma_{k+1}}
\bigg[
\ln\alpha^b_{k+1}(\sigma_{k+1})
+\ln\tilde{\mathcal{J}}(a_k,\sigma_k)
+\ln p(a_k)
\bigg].\label{eqn:logbackwardmsgupdate}
\end{align}
In each maximization, only pairs satisfying
$T(\sigma_k,a_k,\sigma_{k+1})=1$ are considered.

The max-log output used for the symbol decision is the unnormalized
log-posterior score calculated as
\begin{align}
\ln p(a_k|\bm{y})
&\doteq q_{J,k}(a_k;\bm y) \notag\\
&=
\ln p(a_k)
+
\max_{\sigma_k,\sigma_{k+1}}
\bigg[
\ln\alpha^f_k(\sigma_k)+\ln\alpha^b_{k+1}(\sigma_{k+1})
\notag\\&
\qquad\qquad\qquad\qquad\qquad
+\ln\tilde{\mathcal{J}}(a_k,\sigma_k)
\bigg],
\label{eqn:outputdecodermodify}
\end{align}
where the maximization is restricted to valid transitions and we use $q_{J,k}(a_k;\bm y)$ to denote the output of the reduced-state receiver of state length $J$. The receiver
then chooses the symbol with the largest score.

In the absence of a cyclic prefix or suffix, the initial state is fixed by
the boundary condition. In the linear domain, the forward and backward
messages can be initialized as
\[
    \alpha^f(\sigma_0)=1,
    \qquad
    \alpha^b(\sigma_N)=1
\]
for the allowed boundary states. Equivalently, their initial values are zero
in the log domain. 
The forward recursion supplies survivor-path information
about the discarded past, whereas the backward recursion supplies tentative
decisions about the future. Their interaction allows the reduced-state receiver
to approximate the full-state Ungerboeck metric with only $M^J$ states instead
of $M^L$ states. When $J=L$, no past symbols are discarded, and the bias term vanishes. However,
bidirectional bias compensation is specifically required when $J<L$.
In the computation of the forward and the backward messages, the future input is needed, but initially this information is missing. Thus, we do not include bias correction in the initial forward and backward computations. In other words, when making a decision in \eqref{eqn:decisionbias}, the bias term $B_k\left(
\{\dot a_i\}_{i=k-L+1}^{k-J},\{\hat a_i\}_{i=k+1}^{k+L-J}
\right)$ is not subtracted. However, the traceback terms $-\sum_{m=J+1}^{L}a_{k-m}^*\dot a_kg_m$ will still be included as an estimation for the true likelihood metric $\mathcal J (a_k,\sigma_{k-1})$.  We call the completion of the forward and backward process together as one iteration. In order to increase the accuracy of the decision making, it is a good practice to increase the number of iterations. 
 Additional iterations refine the
future-symbol estimates and the bias correction. As will be shown in Section~\ref{sec:simulation}, in our experiments, two
iterations are generally sufficient to obtain a stable performance, especially
under strong ISI.

\section{Error Event Analysis }

% In this section, we analyze how the reduced-state
% receiver can select an incorrect trellis path. We compare the correct
%  path with a competing path that diverges from a certain point and later
% remerges. The resulting metric difference is then simplified to 
% three components, the RKHS distance between the two paths, the
% interaction with the future-symbol interference, and the contribution
% of the filtered noise. This decomposition shows how state reduction
% and imperfect future-symbol decisions affect the error event and
% allows the corresponding pairwise error probability to be obtained.

% In FTN signaling, intentional ISI couples each symbol
% to both past and future symbols. The reduced-state receiver does not contain
% the complete symbol history in its trellis; instead, it uses traceback
% decisions and tentative future-symbol decisions to approximate the discarded
% interference. These approximations can change the path metrics and cause the
% receiver to select an incorrect trellis path.

% An error-event analysis is therefore necessary to explain how FTN time
% packing, state reduction, and imperfect future-symbol decisions produce
% detection errors, rather than only reporting their BER or AIR consequences.
% To this end, we compare the correct path with a competing path that diverges
% at one point and later remerges. We decompose the resulting metric
% difference into three components: the RKHS distance between the two paths,
% the interaction with the residual future-symbol interference, and the
% contribution of the filtered noise. 

To understand the FTN AIR losses associated with reduced-state processing, we examine how residual interference changes the relative metrics of competing FTN sequences. Traceback and tentative future-symbol decisions approximate interactions omitted from the reduced state and can cause an incorrect path to be favored. We compare the transmitted path with a competing path that diverges and later remerges, separating their metric difference into the RKHS distance, residual future-symbol interference, and filtered noise. This geometric analysis explains how the path metrics become unreliable.

An error happens if the accumulated metric of the wrong path exceeds that of
the correct path. Let $\bar{\sigma}_k$ denote the state sequence of the
correct path and let $\sigma_k$ denote the state sequence of a competing path. The two paths share the state at time $k_1$, diverge after
$k_1$, and remerge at time $k_2$. Therefore,    $\sigma_{k_1}=\bar{\sigma}_{k_1}, 
    \sigma_{k_2}=\bar{\sigma}_{k_2}, 
    \sigma_k\neq\bar{\sigma}_k, k_1<k<k_2$. 
Since $a_k$ labels the transition from $\sigma_k$ to
$\sigma_{k+1}$, an error event occurs when
\begin{align}
    \sum_{k=k_1}^{k_2-1}
    \tilde{\mathcal J}(a_k,\sigma_k)
    >
    \sum_{k=k_1}^{k_2-1}
    \tilde{\mathcal J}(\bar a_k,\bar{\sigma}_k),
    \label{eqn:error_metric_condition}
\end{align}
where $a_k$ denotes the symbol on the wrong path and $\tilde{\mathcal J}(a_k,\sigma_k)$ is defined in \eqref{eqn:reducestatemetric}.
Because the two paths remerge at $\sigma_{k_2}$, the last $J$ symbols
before $\sigma_{k_2}$ are common to both paths. Hence, the symbols that may
differ are indexed from $k_1$ to $k_2-J-1$. Define
\begin{align}
    n_e&\triangleq k_2-k_1-J,\\
    \bar{\bm a}_e
    &\triangleq
    [\bar a_{k_1},\bar a_{k_1+1},\dots,\bar a_{k_2-J-1}]^T,\\
    \bm a_e
    &\triangleq
    [a_{k_1},a_{k_1+1},\dots,a_{k_2-J-1}]^T,\\
    \bm e
    &\triangleq
    \bar{\bm a}_e-\bm a_e.
\end{align}

The future symbols that contribute to the untreated future-symbol
interference are
\begin{align}
    \bar{\bm a}_f
    \triangleq
    [\bar a_{k_2},\bar a_{k_2+1},\dots,
    \bar a_{k_2+L-J-1}]^T.
\end{align}
For the practical bidirectional receiver, the corresponding tentative future
decisions are
\begin{align}
    \hat{\bm a}_f
    \triangleq
    [\hat a_{k_2},\hat a_{k_2+1},\dots,
    \hat a_{k_2+L-J-1}]^T.
\end{align}
We define the future-symbol decision error as
\begin{align}
    \bm\epsilon_f
    \triangleq
    \bar{\bm a}_f-\hat{\bm a}_f.
    \label{eqn:future_decision_error}
\end{align}
When no bias correction is applied, $\hat{\bm a}_f=\bm{0}$ and
$\bm\epsilon_f=\bar{\bm a}_f$. When the future symbols are known perfectly,
$\hat{\bm a}_f=\bar{\bm a}_f$ and $\bm\epsilon_f=\bm{0}$.

The corresponding noise segment is
\begin{align}
    \bm\omega
    \triangleq
    [\eta_{k_1},\eta_{k_1+1},\dots,
    \eta_{k_2-J-1}]^T.
\end{align}

Let $\bm \Psi$ denote the principal submatrix of $\bm G$ associated with the
error interval, and let $\bm \Phi$ denote the submatrix describing the
coupling between the error interval and the future symbols. Their entries are
\begin{align}
    (\bm \Psi)_{r,s}
    &\triangleq g_{r-s},
    \qquad r,s=0,\dots,n_e-1,\\
    (\bm \Phi)_{r,l}
    &\triangleq
    g_{(k_1+r)-(k_2+l)}, \notag \\
    &\qquad
    r=0,\dots,n_e-1,\quad
    l=0,\dots,L-J-1.
\end{align}

\begin{corollary}\label{cor:biasderi}
    For the reduced-state receiver with bidirectional decision feedback, an
    error event occurs when
    \begin{align}
        {\bm e}^{\dagger}\bm\Psi{\bm e}
        +2\text{Re}\left\{
        {\bm e}^{\dagger}\bm\Phi\bm\epsilon_f
        \right\}
        +2\text{Re}\left\{
        {\bm e}^{\dagger}\bm\omega
        \right\}
        <0.
        \label{eqn:condierr}
    \end{align}
    In the absence of bias correction, $\bm\epsilon_f=\bar{\bm a}_f$, and
    \eqref{eqn:condierr} becomes
    \begin{align}
        {\bm e}^{\dagger}\bm\Psi{\bm e}
        +2\text{Re}\left\{
        {\bm e}^{\dagger}\bm\Phi\bar{\bm a}_f
        \right\}
        +2\text{Re}\left\{
        {\bm e}^{\dagger}\bm\omega
        \right\}
        <0.
        \label{eqn:condierr_nobias}
    \end{align}
\end{corollary}

\begin{proof}
    See Appendix~\ref{app:proofforerrorevent} for the detailed proof.
\end{proof}
Equations \eqref{eqn:condierr} and \eqref{eqn:condierr_nobias} say that the competing path differs from the transmitted path over a finite
error interval. The term $\bm e^\dagger\bm{\Psi}\bm e$ represents
the distance between the two corresponding noiseless signal vectors
and favors the correct path. The future-symbol interference $2\text{Re}\left\{
        {\bm e}^{\dagger}\bm\Phi\bm\epsilon_f
        \right\}$ and the
filtered-noise term $2\text{Re}\left\{
        {\bm e}^{\dagger}\bm\omega
        \right\}$ are projections onto the direction connecting the two
paths and reduce this distance. An error occurs when their combined
projection is sufficiently negative to overcome half of the distance $\bm e^\dagger\bm{\Psi}\bm e$. 
Since $\bm\omega$ is the corresponding subvector of the filtered noise,
$\bm\omega\sim\mathcal{CN}
    \left(\bm{0},N_0\bm\Psi\right)$.
Therefore,
$2\text{Re}\{{\bm e}^{\dagger}\bm\omega\}$ is a real Gaussian random
variable with zero mean and variance, we have
\begin{align}
    \operatorname{var}
    \left(
    2\text{Re}\left\{{\bm e}^{\dagger}\bm\omega\right\}
    \right)
    =
    2N_0{\bm e}^{\dagger}\bm\Psi{\bm e}.
\end{align}
The probability of error for one particular error event is consequently 
\begin{align}
    P_e(\bm{\sigma}|\bar{\bm{\sigma}})
    =
    Q\left(
    \frac{
    {\bm e}^{\dagger}\bm\Psi{\bm e}
    +2\text{Re}\left\{
    {\bm e}^{\dagger}\bm\Phi\bm\epsilon_f
    \right\}
    }
    {
    \sqrt{
    2N_0{\bm e}^{\dagger}\bm\Psi{\bm e}
    }
    }
    \right).
    \label{eqn:Peoneerrorevent}
\end{align}

The following derivations rewrite each term in \eqref{eqn:condierr} as
an RKHS distance or projection, explaining this error mechanism explicitly.
The term ${\bm e}^{\dagger}\bm\Psi{\bm e}$ can be interpreted as a distance
in the RKHS. To express this distance using the full vector space, we
zero-fill $\bm e$ as
\begin{align}
    \tilde{\bm e}
    =
    [0,\dots,0,\bm e^T,0,\dots,0]^T
    \in\mathbb{C}^{N\times 1},
\end{align}
where the nonzero entries occupy positions $k_1$ through $k_2-J-1$.
We have the first item in the numerator in \eqref{eqn:Peoneerrorevent} becomes
\begin{align}
    {\bm e}^{\dagger}\bm\Psi{\bm e}
    &=
    \tilde{\bm e}^{\dagger}\bm G\tilde{\bm e} =
    \tilde{\bm e}^{\dagger}
    \bm G\bm G^{\#}\bm G
    \tilde{\bm e} =
    \left\langle
    \bm G\tilde{\bm e},
    \bm G\tilde{\bm e}
    \right\rangle_{\mathcal H} =
    \lVert\bm G\tilde{\bm e}\rVert_{\mathcal H}^2.
    \label{eqn:normrkhsGe}
\end{align}
Thus, the distance term ${\bm e}^{\dagger}\bm\Psi{\bm e}$ is the squared
RKHS distance between the two corresponding noiseless signal vectors $\bm{Ga}$ and $\bm{G}\bar{\bm{a}}$.
Similarly, we zero-fill the future-symbol decision error as
\begin{align}
    \tilde{\bm\epsilon}_f
    =
    [0,\dots,0,\bm\epsilon_f^T,0,\dots,0]^T
    \in\mathbb{C}^{N\times 1},
\end{align}
where the nonzero entries occupy positions $k_2$ through
$k_2+L-J-1$. Then, the second item in the numerator in \eqref{eqn:Peoneerrorevent} becomes
\begin{align}
    2\text{Re}\left\{
    {\bm e}^{\dagger}\bm\Phi\bm\epsilon_f
    \right\}
    &=
    2\text{Re}\left\{
    \tilde{\bm e}^{\dagger}\bm G\tilde{\bm\epsilon}_f
    \right\} \notag\\
    &=
    2\text{Re}\left\{
    \tilde{\bm e}^{\dagger}
    \bm G\bm G^{\#}\bm G
    \tilde{\bm\epsilon}_f
    \right\} \notag\\
    &=
    2\text{Re}\left\{
    \left\langle
    \bm G\tilde{\bm e},
    \bm G\tilde{\bm\epsilon}_f
    \right\rangle_{\mathcal H}
    \right\}. \label{eqn:73}
\end{align}
For two vectors $\bm u,\bm v\in\mathcal H$ with
$\lVert\bm v\rVert_{\mathcal H}\neq 0$, defining the real RKHS projection as
\begin{align}
    \text{proj}^{\mathbb R}_{\mathcal H}
    (\bm u\rightarrow\bm v)
    \triangleq
    \frac{
    \text{Re}\left\{
    \left\langle\bm v,\bm u\right\rangle_{\mathcal H}
    \right\}
    }
    {
    \lVert\bm v\rVert_{\mathcal H}
    }.
\end{align}
We can further write \eqref{eqn:73} as 
\begin{align}
    2\text{Re}\left\{
    {\bm e}^{\dagger}\bm\Phi\bm\epsilon_f
    \right\}
    =
    2\lVert\bm G\tilde{\bm e}\rVert_{\mathcal H}
    \cdot
    \text{proj}^{\mathbb R}_{\mathcal H}
    \left(
    \bm G\tilde{\bm\epsilon}_f
    \rightarrow
    \bm G\tilde{\bm e}
    \right). \label{eqn:75}
\end{align}

Inserting  \eqref{eqn:normrkhsGe} and \eqref{eqn:75} into \eqref{eqn:Peoneerrorevent}, the error probability
 becomes
\begin{align}
    P_e(\bm{\sigma}|\bar{\bm{\sigma}})
    &=
    Q\left(
    \frac{
    \lVert\bm G\tilde{\bm e}\rVert_{\mathcal H}^2
    +2\lVert\bm G\tilde{\bm e}\rVert_{\mathcal H}
    \text{proj}^{\mathbb R}_{\mathcal H}
    \left(
    \bm G\tilde{\bm\epsilon}_f
    \rightarrow
    \bm G\tilde{\bm e}
    \right)
    }
    {
    \sqrt{2N_0}\,
    \lVert\bm G\tilde{\bm e}\rVert_{\mathcal H}
    }
    \right) \notag\\
    &=
    Q\left(
    \frac{
    \lVert\bm G\tilde{\bm e}\rVert_{\mathcal H}/2
    +
    \text{proj}^{\mathbb R}_{\mathcal H}
    \left(
    \bm G\tilde{\bm\epsilon}_f
    \rightarrow
    \bm G\tilde{\bm e}
    \right)
    }
    {\sqrt{N_0/2}}
    \right).
\end{align}

Since $\bm\eta$ belongs to $\mathcal H$ almost surely, its projection onto
$\mathcal H$ is itself and we write
\begin{align}
    \bm G\bm G^{\#}\bm\eta=\bm\eta.
\end{align}
The noise term can therefore be written as
\begin{align}
    2\text{Re}\left\{
    {\bm e}^{\dagger}\bm\omega
    \right\}
    &=
    2\text{Re}\left\{
    \tilde{\bm e}^{\dagger}\bm\eta
    \right\} \notag\\
    &=
    2\text{Re}\left\{
    \tilde{\bm e}^{\dagger}
    \bm G\bm G^{\#}\bm\eta
    \right\} \notag\\
    &=
    2\text{Re}\left\{
    \left\langle
    \bm G\tilde{\bm e},
    \bm\eta
    \right\rangle_{\mathcal H}
    \right\} \notag\\
    &=
    2\lVert\bm G\tilde{\bm e}\rVert_{\mathcal H}
    \cdot
    \text{proj}^{\mathbb R}_{\mathcal H}
    \left(
    \bm\eta\rightarrow\bm G\tilde{\bm e}
    \right).
\end{align}
Consequently, the error condition in \eqref{eqn:condierr} can be rewritten as
\begin{align}
    &\lVert\bm G\tilde{\bm e}\rVert_{\mathcal H}^2
    +2\lVert\bm G\tilde{\bm e}\rVert_{\mathcal H}
    \cdot
    \text{proj}^{\mathbb R}_{\mathcal H}
    \left(
    \bm G\tilde{\bm\epsilon}_f
    \rightarrow
    \bm G\tilde{\bm e}
    \right)
    \notag\\
    &\quad
    +2\lVert\bm G\tilde{\bm e}\rVert_{\mathcal H}
    \cdot
    \text{proj}^{\mathbb R}_{\mathcal H}
    \left(
    \bm\eta\rightarrow\bm G\tilde{\bm e}
    \right)
    <0.
\end{align}
Equivalently,
\begin{align}
    &\frac{\lVert\bm G\tilde{\bm e}\rVert_{\mathcal H}}{2}
    +
    \text{proj}^{\mathbb R}_{\mathcal H}
    \left(
    \bm G\tilde{\bm\epsilon}_f
    \rightarrow
    \bm G\tilde{\bm e}
    \right)
    +
    \text{proj}^{\mathbb R}_{\mathcal H}
    \left(
    \bm\eta\rightarrow\bm G\tilde{\bm e}
    \right)
    <0.
\end{align}

\begin{figure}
    \centering
    \includegraphics[width=0.9\linewidth]{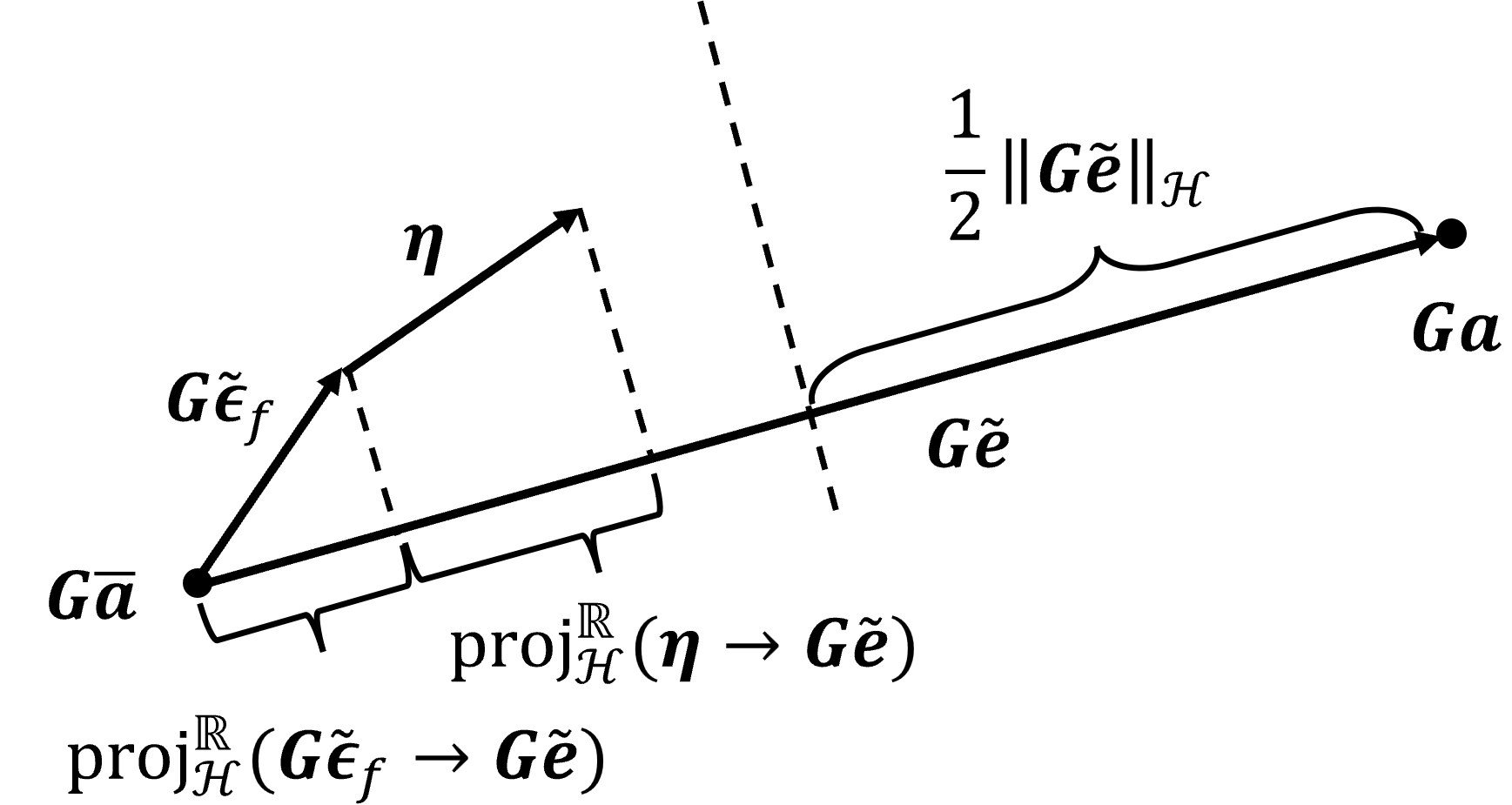}
    \caption{Geometric interpretation of an error event under RKHS
    $\mathcal H$.}
    \label{fig:rkhsfig}
\end{figure}

This can be illustrated in Fig.~\ref{fig:rkhsfig}. It means that an error
event happens if the noise projection is sufficiently large to deviate 
one-half of the relevant RKHS distance together with the residual
future-symbol projection onto the error vector. 

\section{FTN AIR with Reduced-State Detection}
\label{sec:ftn_air_outage}

The preceding sections established the validity of the Ungerboeck
metric under singular FTN noise and explained how state reduction
and future-symbol feedback affect detection errors. We now examine
how these effects influence the information rate supported by the
receiver. 
In FTN signaling, reducing $\delta$ increases the symbol
rate but makes reliable detection more demanding for a receiver
with a fixed state length. Imperfect traceback and future-symbol
decisions can cause the receiver to assign high scores to incorrect
symbols, reducing the information conveyed by its soft outputs.

% \subsection{From Cumulative Path Metrics to Final Symbol Metrics}
% \label{subsec:ftn_air_final_metrics}
% Let $q_{J,k}(c;\bm y)>0$ denote the final score assigned by a
% receiver of state length $J$ to the candidate symbol
% $c\in\mathcal A$ at position $k$. This notation includes the
% full-state receiver with $J=L$ and the reduced-state receiver
% with $J<L$, described in
% Sections~\ref{sec:decodersecAfullstate}
% and~\ref{sec:decodersecBredstate}, respectively.
% Here, $q_{J,k}$ is a linear-domain score; a log-domain
% implementation produces $\ln q_{J,k}$. 
% Define the normalized
% receiver output as
% \begin{equation}
%  p_{J,k}(c\mid\bm y)
%  \triangleq
%  \frac{q_{J,k}(c;\bm y)}
%       {\sum_{d\in\mathcal A}q_{J,k}(d;\bm y)}.
%  \label{eq:ftn_air_normalized_metric}
% \end{equation}
To express the AIR in terms of the final symbol scores,
define the scaled log score as
\begin{equation}
 \Gamma_{J,k}(c;\bm y)
 \triangleq N_0\ln q_{J,k}(c;\bm y).
 \label{eq:ftn_air_symbol_score}
\end{equation}
For the exact full-state receiver,
\eqref{eq:ftn_air_exact_metric} gives
\begin{align}
 \Gamma_{L,k}(c;\bm y)
 &\doteq N_0\ln
 \sum_{\bm a\in\mathcal A_k(c)}
 \exp\!\left(
 -\frac{\|\bm y-\bm G\bm a\|_{\mathcal H}^{2}}{N_0}
 \right),
 \label{eq:ftn_air_score_bridge}
\end{align}
where $\doteq$ denotes equality up to an additive term
independent of $c$. Such a term cancels from score differences
and does not affect the FTN AIR. This expression connects the
RKHS distances of complete candidate sequences to the final
score of a symbol hypothesis. For the reduced-state receiver,
$\Gamma_{J,k}$ is obtained from its actual output after the
prescribed recursions, survivor selections, and feedback
iterations.

\subsection{Exact FTN AIR Decomposition and the Outage Criterion}
\label{subsec:ftn_air_decomposition}

At each symbol position, define
\begin{align}
 \gamma_{J,k}
 &\triangleq
   \max_{c\in\mathcal A}\{\Gamma_{J,k}(c;\bm y)\}
      -\Gamma_{J,k}(\bar a_k;\bm y),
 \label{eq:ftn_air_gamma}\\
 h_{J,k}
 &\triangleq
 \ln\sum_{c\in\mathcal A}
 \exp\!\left(
 \frac{\Gamma_{J,k}(c;\bm y)
 -\max_{d\in\mathcal A}\{\Gamma_{J,k}(d;\bm y)\}}
 {N_0}\right).
 \label{eq:ftn_air_h}
\end{align}
Here, the quantity $\gamma_{J,k}$ is the metric score advantage of
the optimal symbol decision over the transmitted symbol. 
It is zero when 
\begin{align}
    \bar a_k = \underset{c\in \mathcal A}{\arg \max}~\Gamma_{J,k}(c;\bm y),
\end{align}
 and positive when a wrong symbol has a 
strictly greater score.
The quantity $h_{J,k}$ is the log-sum-exp correction term. It
measures the aggregate contribution of all symbol hypotheses relative to
the maximum final metric score. 
Since the maximum includes $\bar a_k$,
\begin{equation}
 \gamma_{J,k}\ge0.
 \label{eq:ftn_air_gamma_nonnegative}
\end{equation}
By \eqref{eqn:maprule}, \eqref{eqn:posteriorandscorefull}, \eqref{eqn:outputdecodermodify} and \eqref{eq:ftn_air_symbol_score}, we know that the receiver decision $\hat a_k$ has the largest estimated posterior $p(\hat a_k|\bm y)$, which is proportional to the score $q_{J,k}(\hat a_k;\bm y)$.
By \eqref{eq:ftn_air_symbol_score},  $\Gamma_{J,k}=N_0\ln q_{J,k}$ by  is a
strictly increasing function of $q_{J,k}$, and by \eqref{eqn:maprule}, the decision $\hat a_k$ maximizes
$q_{J,k}$, we have
\begin{align}
    \max_{c\in\mathcal A}\{\Gamma_{J,k}(c;\bm y)\}=\Gamma_{J,k}(\hat a_k;\bm y).
\end{align}
In \eqref{eq:ftn_air_h}, every exponential term is at most one and 
at least one exponential term equals one. Hence
\begin{align}
 1\le&
 \sum_{c\in\mathcal A}
 e^{[\Gamma_{J,k}(c;\bm y)-\max_d\Gamma_{J,k}(d;\bm y)]
 /N_0}
 \le M \notag\\
 \rightarrow&
 0\le h_{J,k}\le\ln M.
 \label{eq:ftn_air_h_bounds}
\end{align}
The logarithmic term $h_{J,k}$ contains the contribution
of the other hypotheses relative to the largest score.

Taking the largest exponential term out of the symbol sum $\sum_{c\in\mathcal A}q_{J,k}(c;\bm y)$ gives
\begin{align}
 \sum_{c\in\mathcal A}q_{J,k}(c;\bm y)
 &=
 e^{\max_d\frac{\Gamma_{J,k}(d;\bm y)}{N_0}}
 e^{h_{J,k}},\\
 \ln\sum_{c\in\mathcal A}q_{J,k}(c;\bm y)
 &=
 \frac{\max_d\Gamma_{J,k}(d;\bm y)}{N_0}
 +h_{J,k}.
 \label{eq:ftn_air_factor_max}
\end{align}
Therefore the sample contribution to the FTN AIR is
\begin{align}
 &\log_2M+\log_2\frac{ q_{J,k}(\bar a_k;\bm y)}
                    {\sum_cq_{J,k}(c;\bm y)}
 \nonumber\\
 &\quad=
 \log_2M+
 \frac{\Gamma_{J,k}(\bar a_k;\bm y)
       -\max_c\Gamma_{J,k}(c;\bm y)}
      {N_0\ln2}
 -\frac{h_{J,k}}{\ln2}
 \nonumber\\
 &\quad=
 \log_2M-\frac{\gamma_{J,k}/N_0+h_{J,k}}{\ln2}.
 \label{eq:ftn_air_sample_decomposition}
\end{align}
Taking expectations and averaging over $k$, we get 
\begin{equation}
 \text{AIR}_{J}(\delta,N_0)
 =\log_2M-\frac{1}{N\ln2}
  \sum_{k=1}^{N}
  \mathbb E\!\left[\frac{\gamma_{J,k}}{N_0}+h_{J,k}\right].
 \label{eq:ftn_air_exact_decomposition}
\end{equation}
% Note that the computed AIR can have a negative value, but a negative rate does not make any sense. This is caused by the receiver mistakenly assigning too much confidence on the wrong symbol. In this case, the output of the decoder is a wrong decision and the transmission fails.  Therefore, we call this situation as outage, and we regulate the AIR in this case as zero.
The expression in \eqref{eq:ftn_air_exact_decomposition} can be nonpositive when the penalties associated with incorrect confidence and soft ambiguity outweigh the constellation entropy. Its nonnegative part then equals zero, which we refer to as zero-AIR outage. This characterizes the rate obtained from the specified receiver metric for the chosen FTN parameters.
% Multiplying this inequality by the positive number
% $\sigma_0^2\ln2$ gives \eqref{eq:ftn_air_outage_iff}.
Consequently,
\begin{align}
 \text{AIR}_{J}(\delta,N_0)^+=0
 \quad\Longleftrightarrow\quad
 \frac1N\sum_{k=1}^{N}
 \mathbb E[\frac{\gamma_{J,k}}{N_0}+h_{J,k}]
 \ge\ln M.
 \label{eq:ftn_air_outage_iff}
\end{align}
 The quantity $\gamma_{J,k}$ measures the score advantage of
the best candidate over the transmitted symbol. It is zero when the
transmitted symbol has the largest score and becomes positive when the
receiver assigns a larger score to an incorrect symbol. Since
$\gamma_{J,k}$ is divided by $N_{0}$, even a small wrong-symbol score
advantage can cause a large AIR penalty at high SNR.
The term $h_{J,k}$ measures the remaining uncertainty among the candidate
symbols after the largest score has been removed. It is small when one
candidate clearly dominates and increases when several candidates have
similar scores. Thus, $\gamma_{J,k}/N_{0}$ represents the penalty caused by
incorrect confidence, whereas $h_{J,k}$ represents the penalty caused by
soft ambiguity.

With $J$ and $N_0$ fixed, the exact
acceleration factor outage set is therefore
\begin{align}
 \mathcal O_{J,N_0}
 =\bigg\{\delta\in(0,1]:
 \frac1N\sum_{k=1}^{N}
 \mathbb E[\gamma_{J,k}+h_{J,k}]
 \ge\ln M\bigg\}.
 \label{eq:ftn_air_outage_set}
\end{align}
To find the critical value of $\delta$ depends on the state length $J$ and SNR. We leave finding the critical $\delta$ value below which the communication goes into outage as future research.

% \begin{remark}
%     At zero noise, an adverse real projection can make
% \eqref{eq:ftn_air_prefix_margin} negative. However,
% Corollary~\ref{cor:ftn_air_margin} requires an advantage in the
% \emph{final} scores $\Lambda_{J,k}(c;\bm y)$.
% In a max-log receiver, this is a comparison between the best
% completed paths for the two symbol hypotheses. In a sum-product
% receiver, it is a comparison between their sums of completed-path
% weights; no individual wrong path need dominate that sum.
% Thus a negative prefix margin identifies a mechanism for
% outage, whereas \eqref{eq:ftn_air_outage_iff} measures its
% effect after all completion and feedback operations.
% \end{remark}

\subsection{High-SNR Outage from Persistent Final-Score Errors}
\label{subsec:ftn_air_high_snr}

As can be observed from \eqref{eq:ftn_air_exact_decomposition}, as $N_0$ tends to zero, the term $\frac{\gamma_{J,k}}{N_0}$ grows. We therefore multiply both sides of \eqref{eq:ftn_air_exact_decomposition} with $N_0$ to isolate its contribution to the AIR and obtain 
\begin{align}
 N_0\text{AIR}_{J}(\delta,N_0) 
 &=N_0\log_2M
 \nonumber\\
 &\quad-
 \frac{1}{N\ln2}
 \sum_k\mathbb E[\gamma_{J,k}+N_0h_{J,k}].
 \label{eq:ftn_air_scaled_identity}
\end{align}
First of all \eqref{eq:ftn_air_h_bounds} implies
\begin{equation}
 0\le\frac{N_0}{N}\sum_k\mathbb E[h_{J,k}]
 \le N_0\ln M\longrightarrow0, \text{as} ~N_0\rightarrow 0
 \label{eq:ftn_air_scaled_h_limit}
\end{equation}
Also note that 
\begin{align}
    N_0\log_2M\rightarrow 0 , \text{as} ~N_0\rightarrow 0. \label{eqn:log2Mgotozero}
\end{align}
By inserting \eqref{eq:ftn_air_scaled_h_limit} and \eqref{eqn:log2Mgotozero} into \eqref{eq:ftn_air_scaled_identity}, we have 
\begin{equation}
 \lim_{N_0\rightarrow0}
 N_0 \text{AIR}_{J}(\delta,N_0)
 =-
 \frac{1}{N\ln2}
 \sum_k\mathbb E[\gamma_{J,k}].
 \label{eq:ftn_air_scaled_limit}
\end{equation}
This shows that as $N_0$ tends to zero or SNR increases to infinity, the metric difference between the receiver decision and the true transmitted symbol dominates the other terms in the FTN AIR, and this difference dominates because it is magnified by increasing SNR. 
Because of \eqref{eq:ftn_air_gamma_nonnegative}, we know that 
\begin{align}
    \mathbb E[\gamma_{J,k}]\geq 0, 
\end{align}
and thus 
\begin{align}
    -
 \frac{1}{N\ln2}
 \sum_k\mathbb E[\gamma_{J,k}]\leq 0. 
\end{align}
% This says that for sufficiently small $N_0$, the computed AIR is dominated by a non-positive term propotional to the difference $\gamma_{J,k}$ between the computed likelihood of the decided symbol and the true transmitted symbol. This scaled difference becomes relatively larger compared to other terms in \eqref{eq:ftn_air_exact_decomposition} as SNR increases. Because of the reduced state, the computed likelihood becomes an estimate of the true likelihood, the detector also makes mistakes because of the wrong estimation of the future symbols. Therefore, when subtracting the bias, the bias is not perfectly removed, and there will be residual interference. This interference can lead to higher confidence for the wrong symbol, causing error even if there is no noise. Therefore, we conclude that asymptotically the AIR becomes non-positive as the SNR goes to infinity. 
% This happens because a wrong symbol advantage occurs with positive probability in at least one of the $N$ symbols.
% It depends on the state length $J$ and acceleration factor $\delta$ indicating the correlation level as well. 
If the average expected wrong-symbol score advantage remains bounded away from zero as \(N_0\) decreases, its contribution in \eqref{eq:ftn_air_sample_decomposition} grows without bound and drives the nonnegative AIR to zero. Whether this occurs depends on the packing factor, receiver state length, and residual feedback errors. This identifies a limitation of the considered reduced-state processing, whereas the full-state AIR approaches the constellation entropy under the condition of Theorem~\ref{thm:airfulllogm}.

\section{Simulation Results}
\label{sec:simulation}

This section evaluates the AIR of FTN signaling with finite constellations and examines its dependence on the accelration factor, SNR, constellation, roll-off factor, and receiver state length.  We set $T=1$ and use RRC pulse shaping with roll-off factor $\beta$.
% In this section, we evaluate the BER and AIR performance of the analyzed
% detector. We set the symbol interval to $T=1$. All simulations use an RRC
% pulse $g(t)$ with roll-off factor $\beta$. 
%Because of the high correlation nature of FTN, for sufficiently small $\delta$ and block length $N$, we will see that the delay spread $L$ of the channel is higher than the block length. 

\begin{figure}[t]
    \centering
    \includegraphics[width=0.95\columnwidth]{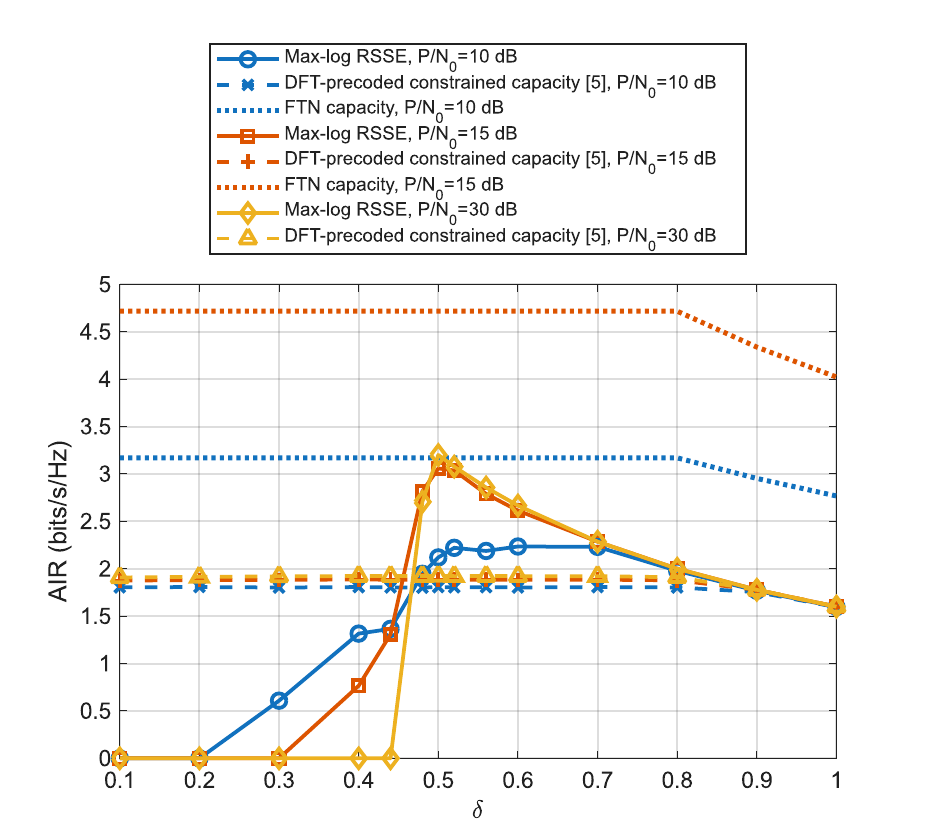}
    \caption{QPSK AIR of RSSE vs DFT-precoded FTN transceiver of  \cite{zhang2026constrainedcapacityanalysisfasterthannyquist} for different $P/N_0$ versus $\delta$, where state length is $J=5$ and traceback length is 45. Roll-off factor for RRC is $\beta=0.25$.  }
    \label{fig:qpsk_ftn_air_vs_tau_rsbcjr_dft_six_curves}
\end{figure}

% We next show the AIR of RSSE computed with \eqref{eq:ftn_air_definition}. 
% In Fig.~\ref{fig:qpsk_ftn_air_vs_tau_rsbcjr_dft_six_curves},
% we compare the AIR of receiver processing, which is RSSE, with transmitter precoding to show the excellence of the proposed detector. 
%  We let the AIR under outage be zero. As can be seen in Fig.~\ref{fig:qpsk_ftn_air_vs_tau_rsbcjr_dft_six_curves}, the AIR increases first as $\delta$ decreases because the acceleration brings in gain in transmission rate. After the transmission accelerates to a certain point, the AIR starts to decrease, this is because as the transmission accelerates, the delay-spread of the channel $L$ increases, the correlation between samples strengthens. This way, the bias term is the correlation of more symbols and the correlation between past and future symbols also becomes stronger, therefore the energy of the bias becomes larger and the detector faces stronger interference. Therefore, after $\delta$ decreases to a certain value, the detector constantly makes a wrong decision and the system is in outage. As SNR increases, the reduced-state decoder assigns too much confidence to the wrong symbol. This effect is magnified by the low SNR as was discussed in Section~\ref{subsec:ftn_air_high_snr}, this is also why we observe that the higher SNR curves go into outage at an earlier stage, namely, larger $\delta$, since the wrong confidence assigned by the detector is magnified. 

Fig.~\ref{fig:qpsk_ftn_air_vs_tau_rsbcjr_dft_six_curves}
compares the achievable spectral efficiency of reduced-state sequence estimation (RSSE) \cite{Gokhanhocampaper} with
that of DFT-precoded FTN
\cite{zhang2026constrainedcapacityanalysisfasterthannyquist}.
For RSSE, we plot
$[\text{AIR}_{J}(\delta,N_0)]^{+}/[\delta(1+\beta)]$,
where $\text{AIR}_{J}$ is given by
\eqref{eq:ftn_air_definition} and nonpositive values are
set to zero. As $\delta$ decreases, the higher symbol
rate initially improves spectral efficiency. However,
more aggressive packing strengthens the ISI and the
coupling between past and future symbols. Residual bias
and imperfect feedback then distort the receiver's
scores, eventually causing outage. At high SNR, the
AIR penalty associated with confidently incorrect
decisions is amplified, as discussed in
Section~\ref{subsec:ftn_air_high_snr}. This explains
the observed range of outages at larger $\delta$
for the higher-SNR curves.

 In Fig.~\ref{fig:qpsk_ftn_air_vs_tau_rsbcjr_dft_six_curves}, we also compare the AIR of RSSE with the discrete-time Fourier transform (DFT)-precoded FTN transceiver presented in \cite{zhang2026constrainedcapacityanalysisfasterthannyquist}. The DFT-precoded FTN transceiver precodes the symbol with the inverse DFT matrix. At the receiver, the received samples are multiplied with the DFT matrix. 
 In the examined configurations, time-domain FTN with reduced-state detection achieves higher spectral efficiency than the DFT-precoded FTN baseline under finite-constellation constraints. 
Fig.~\ref{fig:qpsk_ftn_air_vs_tau_rsbcjr_dft_six_curves} also highlights that precoded transmission is not necessarily
optimal for finite constellation inputs. Although transmission based on eigenmode decomposition
is optimal for Gaussian inputs, this does not generally hold for finite constellation inputs such as QPSK or
QAM. In our results, the time-domain RSSE receiver achieves a higher AIR
than the DFT-precoded scheme, even without transmitter precoding and with
a reduced state.  However, this result does not imply
that the AIR exceeds the unrestricted finite-constellation channel
capacity.
 % This comparison demonstrates the importance of the transmission and detection structure when assessing FTN rates beyond Gaussian-input capacity results.
 %In transmitter-side processing or DFT-precoded FTN, the transceiver uses a total of $2WT$ signaling dimensions  and puts a symbol drawn from a discrete constellation in each signaling dimension, as $\delta$ decreases, the total number of signaling dimension does not change, which is why the AIR saturates after $\delta$ decreases to a certain point, because the essense of DFT-precoded FTN is to fully utilize the unused signaling dimensions in Nyquist transmission. However, in Rx processing FTN, or RSSE, we put more symbols into the $2WT$ resource and each symbol carry meaningful information. This way, we do not send one symbol in each signaling dimension but we utilize the entire $2WT$ dimensions as a whole, and in average in each signaling dimension we are able to send in more information than a discrete symbol can carry, which means in each signaling dimension we can send information more than $\log_2M$. 
The difference lies in
how the fixed $2WT$ signaling dimensions are used. In DFT-precoded
FTN, one symbol from an $M$-ary constellation is assigned to each
available signaling dimension. Reducing $\delta$ initially allows the
dimensions unused by Nyquist signaling to be exploited. Once all
$2WT$ dimensions are utilized, further reducing $\delta$ cannot create
additional dimensions, and the FTN AIR therefore saturates for DFT-precoded FTN. 
The RSSE receiver uses the same $2WT$ resource differently. It
transmits symbols from an $M$-ary constellation every $\delta T$ seconds and jointly detects the whole sequence. As $\delta$ decreases, more $M$-ary symbols are packed into the same $2WT$ dimensions. They are not assigned
one-to-one to orthogonal dimensions; instead, they form a joint
codeword. If these codewords remain distinguishable, their joint
information can produce an aggregate spectral efficiency exceeding
$\log_2 M$ per normalized dimension, although no individual symbol
carries more than $\log_2 M$ bits. However, under excessive packing, residual
bias reduces distinguishability and drives the RSSE AIR into outage.

\begin{figure}
    \centering
    \includegraphics[width=0.85\columnwidth]{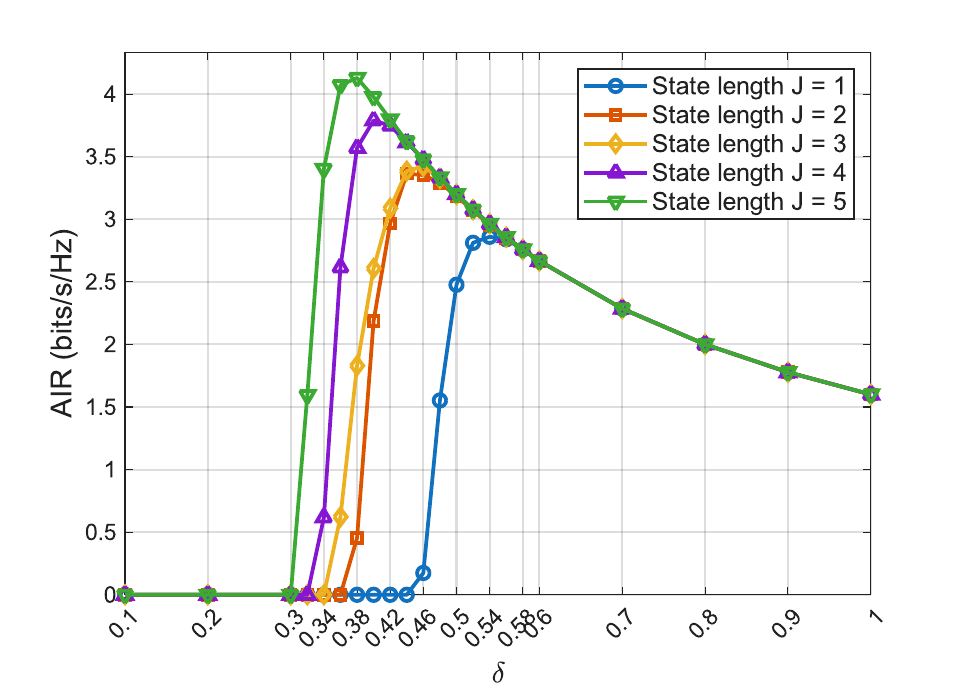}
    \caption{QPSK AIR of RSSE with fixed $P/N_0=20$ dB versus $\delta$ for different state lengths$J$, where the traceback length is 120. Roll-off factor for RRC is $\beta=0.25$. }
    \label{fig:qpsk_ftn_air_vs_tau_state_sweep_20dbtxsnr}
\end{figure}

In Fig.~\ref{fig:qpsk_ftn_air_vs_tau_state_sweep_20dbtxsnr}, we compare the AIR of RSSE for different state lengths. We find that when the state length $J$ is larger, the AIR can reach higher values before it begins to decrease. This is because for larger $J$, the receiver will more likely make a correct decision, and the AIR also goes into outage below a smaller $\delta$ value. As $J$ grows, more symbols are captured and represented by the state, this way less symbols cause interference in the bias term, the bias term has less energy and the receiver assigns less confidence to the wrong symbol.   

\begin{figure}[t]
    \centering
    \includegraphics[width=0.8\columnwidth]{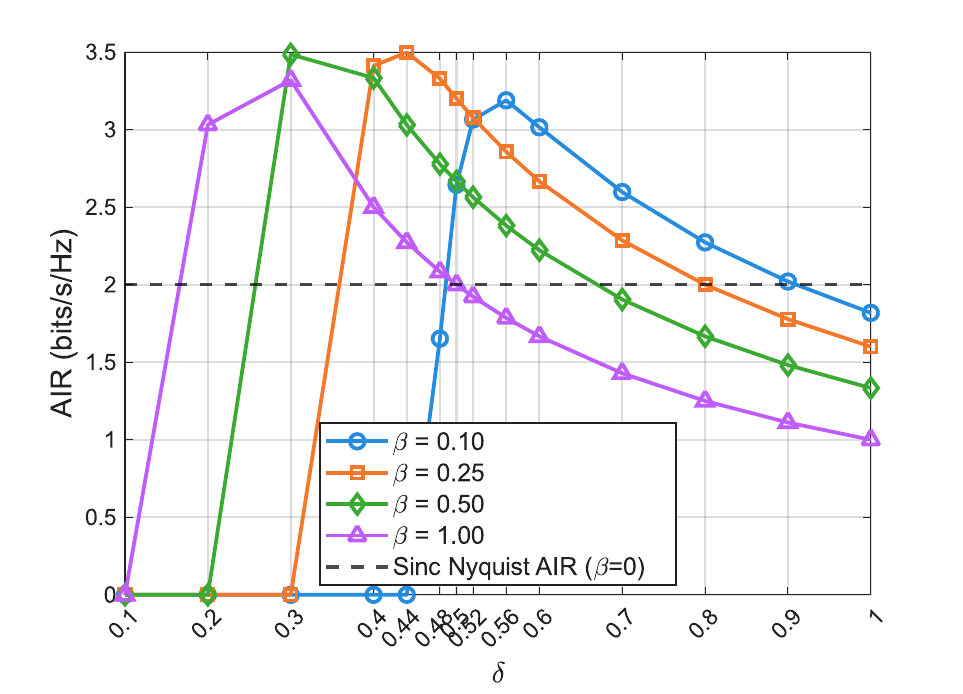}
    \caption{QPSK AIR of RSSE with fixed $P/N_0=20$ dB versus $\delta$ for different roll-off factor $\beta$. The state length $J=5$, and the traceback length is 45. }
    \label{fig:qpsk_ftn_air_vs_tau_beta_sweep_txsnr_addnyq}
\end{figure}

We next show the effect of $\beta$ on the AIR of RSSE in Fig.~\ref{fig:qpsk_ftn_air_vs_tau_beta_sweep_txsnr_addnyq}. As $\beta$ increases, the curve $g(t)$ has weaker sidelobes, and its energy is more concentrated in time; samples  $g(n\delta T)$ for $n\neq0$ also become smaller relative to $g(0)$. Therefore, for fixed $\delta$, if we increase $\beta$, the correlation between symbols, which is indicated by the coefficients $g(n\delta T), n=\pm1, \dots, \pm(N-1)$, becomes weaker. The effective ISI and the energy of the residual bias are consequently reduced. Hence, the AIR-outage transition is
shifted toward smaller $\delta$, which means stronger time
acceleration can be allowed because of the compensation of larger $\beta$.

\begin{figure}[t]
    \centering
    \includegraphics[width=0.8\columnwidth]{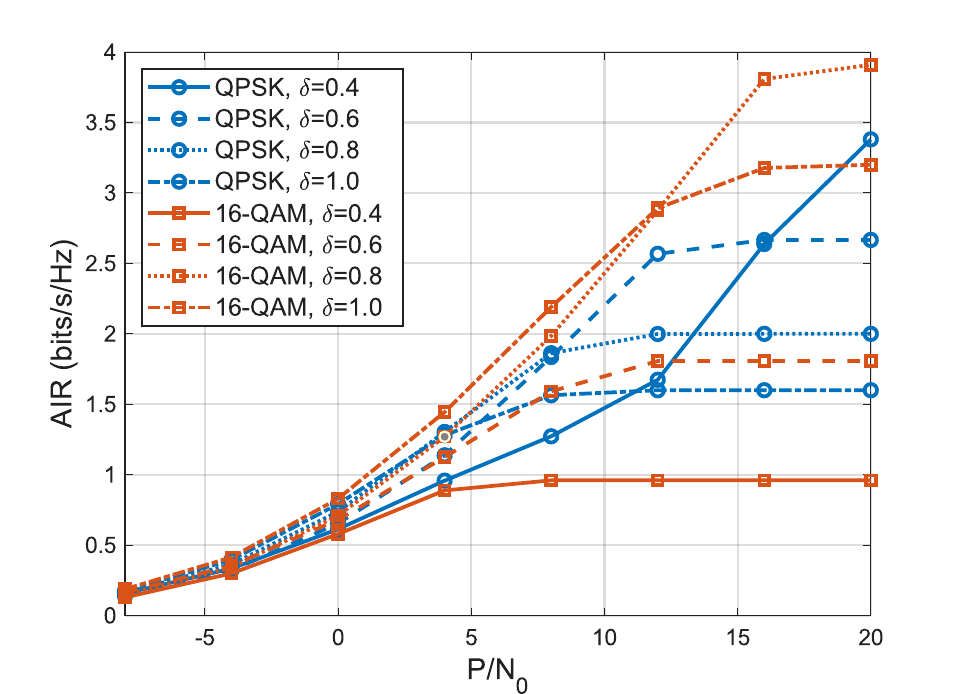}
    \caption{AIR of RSSE versus $P/N_0$ for QPSK and 16 QAM with different $\delta$ values, where the state length $J$ for QPSK is 5 and $J$ for 16 QAM is 3. Traceback length is 45 and roll-off factor $\beta=0.25$. }
    \label{fig:air_vs_snr_qpsk_16qam_tau_sweep_tx}
\end{figure}

% \begin{figure}[t]
%     \centering
%     \includesvg[width=0.8\columnwidth,
%         inkscapelatex=false,
%         inkscapearea=page]{rx_paper.svg}
%     \caption{AIR of RSSE versus $E_s/N_0$ for QPSK and 16 QAM with different $\delta$ values, where the state length $J$ for QPSK is 5 and $J$ for 16 QAM is 3. Traceback length is 45 and roll-off factor $\beta=0.25$.}
%     \label{fig:air_vs_snr_qpsk_16qam_tau_sweep_rx}
% \end{figure}

% In Figs.~\ref{fig:air_vs_snr_qpsk_16qam_tau_sweep_tx} and~\ref{fig:air_vs_snr_qpsk_16qam_tau_sweep_rx}, we plot the AIR of QPSK and 16 QAM constellations for different $\delta$ for fixed $P/N_0$ and  $E_s/N_0$ respectively. We choose smaller state length $J$ for 16 QAM curves for lower receiver complexity. In both figures, the AIR increases and then saturates as SNR increases since a discrete constellation is used, and the information each symbol can carry has a limit. 
% We can see that 16 QAM reaches the decreasing part of the AIR curve at a smaller $\delta$ than QPSK, since 16 QAM constellation has a smaller minimum Euclidean distance compared to QPSK, which is more prone to mistakes and results in a residual bias, which causes interference. 

Fig.~\ref{fig:air_vs_snr_qpsk_16qam_tau_sweep_tx}
shows the AIR of QPSK and 16-QAM versus $P/N_0$ for
different packing factors $\delta$. We use $J=5$ for
QPSK and $J=3$ for 16-QAM, choosing the shorter state
length to limit receiver complexity. The AIR generally
increases with SNR, with several curves approaching a
plateau. At $P/N_0=20$~dB, reducing $\delta$ improves
the QPSK AIR over the values considered, whereas
16-QAM achieves its highest AIR at $\delta=0.8$ and
loses rate with further packing. For these receiver
configurations, the losses under stronger ISI can
therefore outweigh the gain from transmitting more
symbols. Results plotted against $E_s/N_0$ show
similar qualitative trends and are omitted for brevity.

\begin{figure}[!t]
    \centering
    \includegraphics[width=0.85\columnwidth]{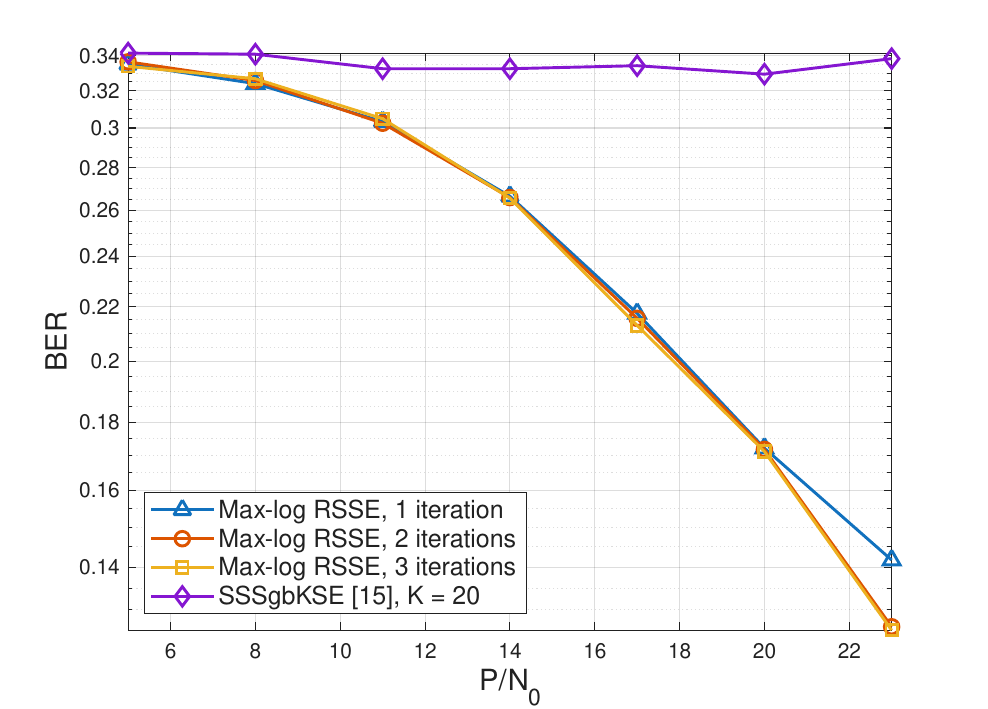}
    \caption{QPSK BER performance of RSSE vs SSSgbKSE of  \cite{lowcomplxsssse}  under fixed $P/N_0$, where $\delta=0.3$ and $\beta=0.25$. Results for RSSE with 1 to 3 iterations are shown.  The state length is 5 and the traceback length $L-J$ is 45.  }
    \label{fig:ber_vs_snr_rsbcjr_2iter_3iter_gbk_k20}
\end{figure}

% In Fig.~\ref{fig:ber_vs_snr_rsbcjr_2iter_3iter_gbk_k20}, we demonstrate the effect of the number of iterations, in other words, the effect of bias correction. We first compare the BER performance of RSSE for 1 to 3 iterations. For RSSE with 1 iteration, there is no bias correction in the decision making, the state path finding process is therefore interfered. However, as we bring in bias correction in the second iteration, the estimation for future symbol is present after the first iteration, which assists with the bias elimination and the decision making, we therefore achieve better performance than the one without bias correction. After we increase the number of iterations to 3, since the survival path construction in the second iteration was well assisted by bias correction, the third iteration further refines the decisions made in the second iteration. However, as can be observed in Fig.~\ref{fig:ber_vs_snr_rsbcjr_2iter_3iter_gbk_k20}, the survival path was already well-constructed in the second iteration so the third iteration brings very slight improvement. We also compare the RSSE performance with SSSgbKSE in \cite{lowcomplxsssse} for $K=20$. We can see that under extreme acceleration, or the correlation is too strong, low-complexity detector is not able to perform well. 

Fig.~\ref{fig:ber_vs_snr_rsbcjr_2iter_3iter_gbk_k20}
demonstrates a substantial BER advantage of RSSE over
SSSgbKSE \cite{lowcomplxsssse} under severe FTN time
packing. 
SSSgbKSE is among the simplest FTN receivers because it performs
successive symbol-by-symbol detection with a finite go-back parameter
\(K\). In contrast, the proposed RSSE provides a tunable
complexity and performance trade-off through its state length \(J\).
When \(J=L\), no state merging is performed and the receiver becomes
the full-state BCJR detector. Its max-log implementation is
equivalent to ML sequence detection, whereas its sum-product
implementation provides the exact symbol posterior. The plotted
\(J=5\) RSSE receiver therefore represents an intermediate-complexity
operating point between SSSgbKSE and full-state ML detection.
Despite retaining only $J=5$ symbols in the trellis
state,  RSSE
 handles the strong  ISI introduced
by FTN more effectively than SSSgbKSE with
$K=20$. Moreover, RSSE performance demonstrates that it is able to further decrease the BER as SNR keeps increasing, while SSSgbKSE is not able to perform well regardless of SNR value. This is because SSSgbKSE is not able to cancel the ISI properly. 
The iteration comparison illustrates the benefit of
bias correction. The first iteration operates without
bias correction, whereas the second uses tentative
future-symbol decisions from the first iteration to
improve survivor-path selection. This provides an
additional BER reduction, particularly at higher
$P/N_0$. A third iteration produces negligible further
improvement, indicating that two iterations capture
most of the observed benefit.

\begin{figure}[t]
    \centering
    \includegraphics[width=0.9\columnwidth]{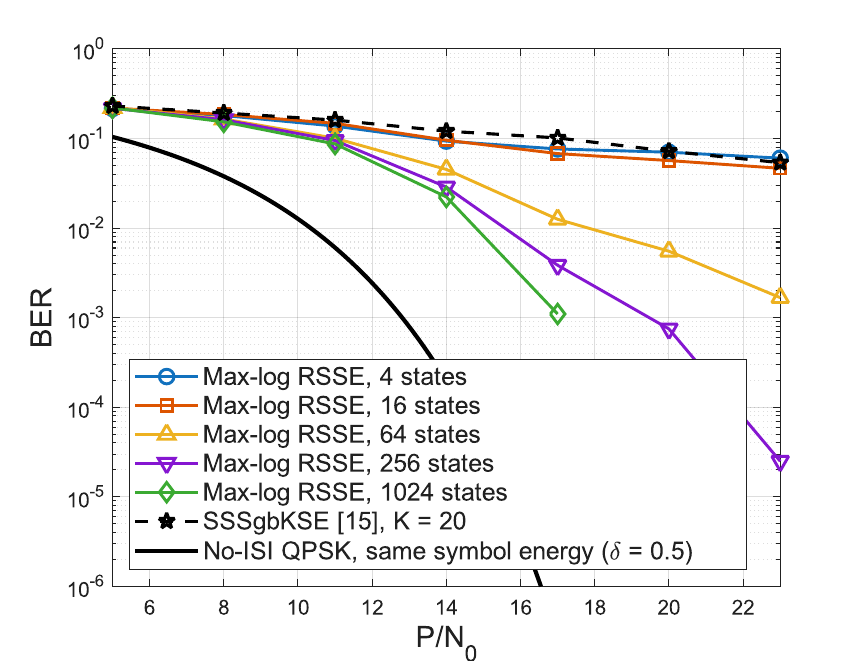}
    \caption{QPSK BER performance of RSSE vs SSSgbKSE of \cite{lowcomplxsssse} with $K=20$ under fixed $P/N_0$, where $\delta=0.5$ and $\beta=0.25$.  Results for RSSE for different state length $J$ for 2 iterations is shown. The total memory length $L=50$. }
    \label{fig:qpsk_ber_bcjr_state_sweep_gbk_txsnr}
\end{figure}

In Fig.~\ref{fig:qpsk_ber_bcjr_state_sweep_gbk_txsnr}, we show the effect of different state lengths. As we increase the state length, we have better estimation for the true likelihood and therefore we can have better BER results. On the other hand, reducing the state length from $L$ to $J$ results in the fact that  the maximum number of paths can be represented by the trellis is reduced to $M^{J}$. Since we need $M^L$ states to represent all $M^L$ paths,  path merging will occur. In this case, the smaller the state length is, the fewer paths we can represent, and it is more likely to trim the correct path in the survivor path construction.

\begin{figure}[t]
    \centering
    \includegraphics[width=0.8\columnwidth]{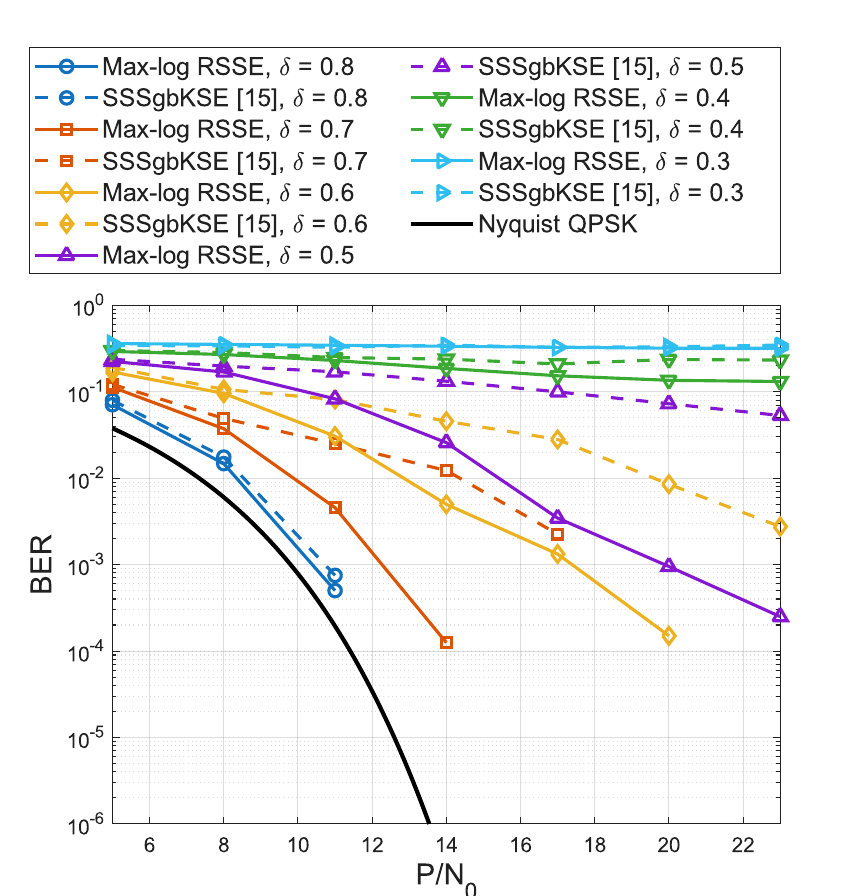}
    \caption{QPSK BER performance of RSSE vs SSSgbKSE of \cite{lowcomplxsssse} with $K=20$ under fixed $P/N_0$, where  $\beta=0.25$.   
     Results for RSSE and SSSgbKSE for different $\delta$ for 2 iterations are shown. The state length is $J=5$, and the traceback length is $L-J=45$. }
    \label{fig:qpsk_ber_bcjr_vs_SNR_delta_sweep}
\end{figure}

In Fig.~\ref{fig:qpsk_ber_bcjr_vs_SNR_delta_sweep}, we demonstrate the effect of $\delta$ on the BER performance of RSSE and compare with the performance of low-complexity detection of SSSgbKSE. As $\delta$ decreases, the correlation between symbols increases and the delay-spread of the channel increases. Therefore, the detection becomes more difficult. Comparing the RSSE results and the SSSgbKSE results, we find that at high $\delta$ values, for example, $\delta=0.8$, the performance of RSSE and SSSgbKSE do not have a big difference. However, as $\delta$ decreases below the critical threshold $\frac{1}{1+\beta}$, the SSSgbKSE performance starts to perform poorly compared to RSSE.

\section{Conclusion}

This paper investigated the achievable information rates and spectral efficiency of FTN signaling with finite constellations, developing full-state and reduced-state BCJR receivers to characterize the gains from time packing and the limitations imposed by receiver complexity. To analyze detection under strong correlation among received symbols, we formulated the FTN detection problem in the reproducing kernel Hilbert space induced by the correlation matrix $\bm{G}$. We showed that the signal and filtered noise lie in the same reproducing kernel Hilbert space almost surely, ensuring that the conventional Ungerboeck branch metric remains well-defined even when $\bm{G}$ is rank deficient.

We also investigated how reducing the receiver state length affects detection performance. Although state reduction lowers computational complexity, it introduces residual interference that can lead to incorrect symbol decisions. We showed that bi-directional feedback can help mitigate this effect through bias correction and provided a geometric interpretation of the role of residual bias in detection errors.

Our results further revealed that more aggressive FTN acceleration does not necessarily translate into higher spectral efficiency when receiver complexity is limited. Under strong ISI, errors in feedback and traceback can lead to incorrect decisions that persist even at high SNR, causing the achievable information rate to deteriorate. In contrast, full-state detection approaches $\log_2 M$ bits per symbol at high SNR, provided that $d_{\min}(\delta)>0$. This shows that the high-SNR rate collapse is not an inherent limitation of FTN signaling, but rather a consequence of receiver approximations.

Overall, our numerical results highlight the importance of jointly considering FTN acceleration, constellation size, SNR, and receiver complexity when designing practical FTN systems. Rather than simply minimizing the acceleration factor, the goal should be to identify operating conditions under which increased transmission rates translate into actual spectral-efficiency gains.

\appendices

\section{Proof for Theorem~\ref{thm:metricworks}}\label{app:metricworks}

In this appendix, we prove Theorem~\ref{thm:metricworks}.
We write the term inside the exponential of  \eqref{eqn:likelihoodbelowthres} as 
\begin{align}
    &-\frac{1}{2}(\bm{y}-\bm{G}\bm{a})^\dagger\tilde{\bm{G}}^{\#}(\bm{y}-\bm{G}\bm{a})\notag\\&=-\frac{1}{2\sigma_0^2}\left(\bm{y}^\dagger\bm{G}^\#\bm{y}-2\text{Re}\left\{\bm{a}^\dagger\bm{G}\bm{G}^\#\bm{y}\right\}+\bm{a}^\dagger\bm{G}\bm{G}^\#\bm{G}\bm{a}\right)\notag\\
    &\overset{(a)}{=}-\frac{1}{2\sigma_0^2}\left(\bm{y}^\dagger\bm{G}^\#\bm{y}-2\text{Re}\left\{\bm{a}^\dagger\bm{G}\bm{G}^\#\bm{y}\right\}+\bm{a}^\dagger\bm{G}\bm{a}\right), \label{eqn:deriv1}
\end{align}
where (a) is because of the property of the Moore-Penrose inverse $\bm{G}\bm{G}^\#\bm{G}=\bm{G}$. 
Notice that in \eqref{eqn:deriv1}, the product $\bm{G}\bm{G}^\#$ is a projection onto the range of $\bm{G}$, we have 
\begin{align}
    \bm{G}\bm{G}^\#\bm{y}=\bm{y}.
\end{align}
Therefore, we continue \eqref{eqn:deriv1} as
% \begin{align}
%     -\frac{1}{2}(\bm{y}-\bm{G}\bm{a})^\dagger\tilde{\bm{G}}^{\#}(\bm{y}-\bm{G}\bm{a})=-\frac{1}{2\sigma_0^2}\left(\bm{y}^\dagger\bm{G}^\#\bm{y}-2\text{Re}\left\{\bm{a}^\dagger\bm{y}\right\}+\bm{a}^\dagger\bm{G}\bm{a}\right). \label{eqn:deriv2}
% \end{align}
\begin{align}
&-\frac{1}{2}
(\bm{y}-\bm{G}\bm{a})^\dagger
\tilde{\bm{G}}^{\#}
(\bm{y}-\bm{G}\bm{a})
\notag\\[-0.3ex]
&\quad=
-\frac{1}{2\sigma_0^2}\biggl(
\bm{y}^\dagger\bm{G}^{\#}\bm{y}
-2\text{Re}\left\{
\bm{a}^\dagger\bm{y}
\right\}
+\bm{a}^\dagger\bm{G}\bm{a}
\biggr).
\label{eqn:deriv2}
\end{align}
By observing \eqref{eqn:deriv2}, we can see that the term $\bm{y}^\dagger\bm{G}^\#\bm{y}$ is irrelevant to the data symbols $\bm{a}$. Therefore, in order to maximize the likelihood $p(\bm{y}|\bm{a})$, we only need to find the optimal $\hat{\bm{a}}$
\begin{align}
    \hat{\bm{a}}=\underset{\bm{a}}{\arg\max}~\frac{1}{2\sigma_0^2}\left(2\text{Re}\left\{\bm{a}^\dagger\bm{y}\right\}-\bm{a}^\dagger\bm{G}\bm{a}\right).
\end{align}
We can further decompose the metric into 
\begin{align}
    p(\bm{y}|\bm{a})&=\sum_{k=1}^{N}J(\sigma_k,\sigma_{k-1}) \notag\\
    &=\sum_{k=1}^{N}2\text{Re}\{a^*_ky_k\}-|a_k|^2g[0]-2\text{Re}\{a_k^*\sum_{l=1}^{L}g[l]a_{k-l}\}. \label{eqn:103}
\end{align}
Equation~\eqref{eqn:103} shows that the sequence-dependent part of the
likelihood factorizes into the product of the branch metrics in
\eqref{eqn:likelihoodbelowthres}. Therefore, the Ungerboeck branch metric remains valid even
when the Gram matrix $\bm{G}$ is singular, which completes the proof of
Theorem~\ref{thm:metricworks}.

\section{Proof for Corollary~\ref{cor:biasderi}}
\label{app:proofforerrorevent}

\begin{figure}[t]
    \centering
    \includegraphics[width=0.9\linewidth]{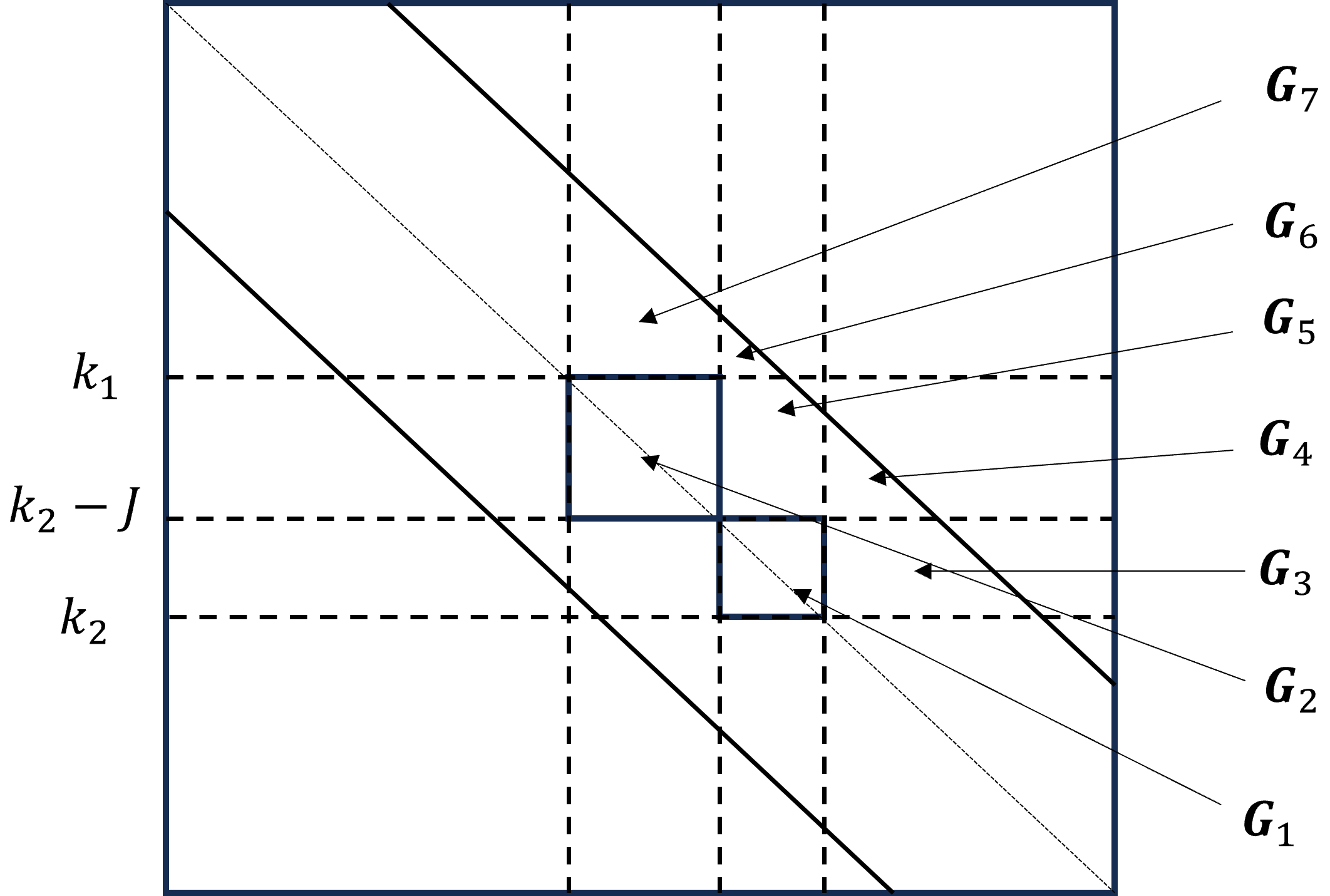}
    \caption{Block partition of the Gram matrix $\bm G$.}
    \label{fig:matlabel}
\end{figure}

In this appendix we prove Corollary~~\ref{cor:biasderi}. Fig.~\ref{fig:matlabel} illustrates the partition of the Toeplitz
Gram matrix associated with the past-symbol, error-event, common-symbol,
and future-symbol intervals.

% \begin{figure}[t]
%     \centering
%     \includegraphics[width=0.9\linewidth]{matlabel.png}
%     \caption{Block partition of the Gram matrix used in the error-event
%     analysis.}
%     \label{fig:matlabel}
% \end{figure}

The two paths diverge after \(k_1\) and remerge at \(k_2\). Therefore,
the symbols that may differ between the two paths are indexed from
\(k_1\) to \(k_2-J-1\). Define
\[
n_e\triangleq k_2-k_1-J,
\qquad
n_f\triangleq L-J .
\]

The four symbol vectors involved in the computation are
\begin{align}
\bm a_p
&=\{\bm a\}_{k_1-L}^{\,k_1-1}
=[a_{k_1-L},a_{k_1-L+1},\ldots,a_{k_1-1}]^T,\\
\bm a_e
&=\{\bm a\}_{k_1}^{\,k_2-J-1}
=[a_{k_1},a_{k_1+1},\ldots,a_{k_2-J-1}]^T,\\
\bm a_J
&=\{\bm a\}_{k_2-J}^{\,k_2-1}
=[a_{k_2-J},a_{k_2-J+1},\ldots,a_{k_2-1}]^T,\\
\bm a_f
&=\{\bm a\}_{k_2}^{\,k_2+L-J-1}
=[a_{k_2},a_{k_2+1},\ldots,a_{k_2+L-J-1}]^T.
\end{align}

The received-symbol vectors associated with the error and common
intervals are respectively given as
\begin{align}
\bm y_e
&=[y_{k_1},y_{k_1+1},\ldots,y_{k_2-J-1}]^T,\\
\bm y_J
&=[y_{k_2-J},y_{k_2-J+1},\ldots,y_{k_2-1}]^T.
\end{align}
The barred versions
\(\bar{\bm a}_p,\bar{\bm a}_e,\bar{\bm a}_J,\bar{\bm a}_f\)
denote the corresponding actual transmitted-symbol vectors. The
future-symbol decisions used by the receiver are
\[
\hat{\bm a}_f
=
[\hat a_{k_2},\hat a_{k_2+1},\ldots,
\hat a_{k_2+L-J-1}]^T,
\]
and the future-symbol decision error is
\[
\bm\epsilon_f
\triangleq
\bar{\bm a}_f-\hat{\bm a}_f .
\]
The noise vectors corresponding to the two observation intervals are
\begin{align}
\bm\eta_e
&=[\eta_{k_1},\eta_{k_1+1},\ldots,\eta_{k_2-J-1}]^T,\\
\bm\eta_J
&=[\eta_{k_2-J},\eta_{k_2-J+1},\ldots,\eta_{k_2-1}]^T.
\end{align}
Thus, \(\bm\eta_e\) is the same filtered-noise vector denoted by
\(\bm\omega\) in the main text:
\[
\bm\eta_e=\bm\omega .
\]
For convenience, define
\begin{align}
    &p_r=k_1-L+r,\qquad
e_s=k_1+s,\qquad \\
&c_t=k_2-J+t,\qquad
f_\ell=k_2+\ell .
\end{align}

The matrices shown in Fig.~\ref{fig:matlabel} are defined as
% \begin{align}
% \bm G_1&\in\mathbb C^{J\times J},
% &
% [\bm G_1]_{t,u}
% &=g[c_t-c_u]
%  =g[t-u],
% \\
% \bm G_2&\in\mathbb C^{n_e\times n_e},
% &
% [\bm G_2]_{s,u}
% &=g[e_s-e_u]
%  =g[s-u],
% \\
% \bm G_3&\in\mathbb C^{J\times n_f},
% &
% [\bm G_3]_{t,\ell}
% &=g[c_t-f_\ell]
%  =g[-J+t-\ell],
% \\
% \bm G_4&\in\mathbb C^{n_e\times n_f},
% &
% [\bm G_4]_{s,\ell}
% &=g[e_s-f_\ell]
%  =g[(k_1+s)-(k_2+\ell)],
% \\
% \bm G_5&\in\mathbb C^{n_e\times J},
% &
% [\bm G_5]_{s,t}
% &=g[e_s-c_t]
%  =g[(k_1+s)-(k_2-J+t)],
% \\
% \bm G_6&\in\mathbb C^{L\times J},
% &
% [\bm G_6]_{r,t}
% &=g[p_r-c_t]
%  =g[(k_1-L+r)-(k_2-J+t)],
% \\
% \bm G_7&\in\mathbb C^{L\times n_e},
% &
% [\bm G_7]_{r,s}
% &=g[p_r-e_s]
%  =g[-L+r-s].
% \end{align}
\begin{align}
\bm G_1&\in\mathbb C^{J\times J},
&
[\bm G_1]_{t,u}
&=g[c_t-c_u]=g[t-u],
\\
\bm G_2&\in\mathbb C^{n_e\times n_e},
&
[\bm G_2]_{s,u}
&=g[e_s-e_u]=g[s-u],
\\
\bm G_3&\in\mathbb C^{J\times n_f},
&
[\bm G_3]_{t,\ell}
&=g[c_t-f_\ell]=g[-J+t-\ell],
\\
\bm G_4&\in\mathbb C^{n_e\times n_f},
&
[\bm G_4]_{s,\ell}
&=g[e_s-f_\ell]
\notag\\
&&&=g[(k_1+s)-(k_2+\ell)],
\\
\bm G_5&\in\mathbb C^{n_e\times J},
&
[\bm G_5]_{s,t}
&=g[e_s-c_t]
\notag\\
&&&=g[(k_1+s)-(k_2-J+t)],
\\
\bm G_6&\in\mathbb C^{L\times J},
&
[\bm G_6]_{r,t}
&=g[p_r-c_t]
\notag\\
&&&=g[(k_1-L+r)-(k_2-J+t)],
\\
\bm G_7&\in\mathbb C^{L\times n_e},
&
[\bm G_7]_{r,s}
&=g[p_r-e_s]=g[-L+r-s].
\end{align}

The index ranges are
\[
\begin{aligned}
&t,u=0,\ldots,J-1,\\
&s,u=0,\ldots,n_e-1,\\
&t=0,\ldots,J-1,\qquad \ell=0,\ldots,n_f-1,\\
&s=0,\ldots,n_e-1,\qquad t=0,\ldots,J-1,\\
&r=0,\ldots,L-1.
\end{aligned}
\]

In accordance with the notation used in the main text, the matrices
\(\bm\Psi\) and \(\bm\Phi\) are
\[
\bm\Psi\triangleq\bm G_2,
\qquad
\bm\Phi\triangleq\bm G_4.
\]

Using the partition in Fig.~\ref{fig:matlabel}, the received vectors can
be written as
\begin{align}
\bm y_e
&=
\bm G_7^\dagger\bar{\bm a}_p
+\bm G_2\bar{\bm a}_e
+\bm G_5\bar{\bm a}_J
+\bm G_4\bar{\bm a}_f
+\bm\eta_e,
\label{eq:appendix_ye}
\\
\bm y_J
&=
\bm G_6^\dagger\bar{\bm a}_p
+\bm G_5^\dagger\bar{\bm a}_e
+\bm G_1\bar{\bm a}_J
+\bm G_3\bar{\bm a}_f
+\bm\eta_J .
\label{eq:appendix_yJ}
\end{align}

Let \(\bar{\mathcal M}\) and \(\mathcal M\) denote the accumulated
metrics of the correct and competing paths, respectively, over the
transitions \(k=k_1,\ldots,k_2-1\). Since the two paths are identical
outside the error interval, their path-dependent metric terms can be
written as
\begin{align}
\bar{\mathcal M}
={}&
2\operatorname{Re}\{
\bar{\bm a}_e^\dagger\bm y_e
\}
-2\operatorname{Re}\{
\bar{\bm a}_e^\dagger
\bm G_7^\dagger\bar{\bm a}_p
\}
-\bar{\bm a}_e^\dagger\bm G_2\bar{\bm a}_e
\nonumber\\
&+
2\operatorname{Re}\{
\bar{\bm a}_J^\dagger\bm y_J
\}
-2\operatorname{Re}\left\{
\bar{\bm a}_J^\dagger
[\bm G_6^\dagger,\bm G_5^\dagger]
\begin{bmatrix}
\bar{\bm a}_p\\
\bar{\bm a}_e
\end{bmatrix}
\right\}
-\bar{\bm a}_J^\dagger\bm G_1\bar{\bm a}_J
\nonumber\\
&-
2\operatorname{Re}\{
\bar{\bm a}_e^\dagger\bm G_4\hat{\bm a}_f
\},
\label{eq:appendix_correct_metric}
\end{align}
and
\begin{align}
\mathcal M
={}&
2\operatorname{Re}\{
\bm a_e^\dagger\bm y_e
\}
-2\operatorname{Re}\{
\bm a_e^\dagger
\bm G_7^\dagger\bar{\bm a}_p
\}
-\bm a_e^\dagger\bm G_2\bm a_e
\nonumber\\
&+
2\operatorname{Re}\{
\bar{\bm a}_J^\dagger\bm y_J
\}
-2\operatorname{Re}\left\{
\bar{\bm a}_J^\dagger
[\bm G_6^\dagger,\bm G_5^\dagger]
\begin{bmatrix}
\bar{\bm a}_p\\
\bm a_e
\end{bmatrix}
\right\}
-\bar{\bm a}_J^\dagger\bm G_1\bar{\bm a}_J
\nonumber\\
&-
2\operatorname{Re}\{
\bm a_e^\dagger\bm G_4\hat{\bm a}_f
\}.
\label{eq:appendix_competing_metric}
\end{align}

Since the paths coincide before the divergence and in the common
interval,
\[
\bm a_p=\bar{\bm a}_p,
\qquad
\bm a_J=\bar{\bm a}_J .
\]
Define the error vector
\[
\bm e\triangleq
\bar{\bm a}_e-\bm a_e .
\]

Substituting \eqref{eq:appendix_ye} and
\eqref{eq:appendix_yJ} into
\eqref{eq:appendix_correct_metric} and
\eqref{eq:appendix_competing_metric}, and then subtracting, gives
\begin{align}
\bar{\mathcal M}-\mathcal M
={}&
\bar{\bm a}_e^\dagger\bm G_2\bar{\bm a}_e
-2\operatorname{Re}\{
\bm a_e^\dagger\bm G_2\bar{\bm a}_e
\}
+\bm a_e^\dagger\bm G_2\bm a_e
\nonumber\\
&+
2\operatorname{Re}\{
\bar{\bm a}_e^\dagger\bm G_4\bar{\bm a}_f
\}
-2\operatorname{Re}\{
\bm a_e^\dagger\bm G_4\bar{\bm a}_f
\}
\nonumber\\
&-
2\operatorname{Re}\{
\bar{\bm a}_e^\dagger\bm G_4\hat{\bm a}_f
\}
+2\operatorname{Re}\{
\bm a_e^\dagger\bm G_4\hat{\bm a}_f
\}
\nonumber\\
&+
2\operatorname{Re}\{
\bar{\bm a}_e^\dagger\bm\eta_e
\}
-2\operatorname{Re}\{
\bm a_e^\dagger\bm\eta_e
\}.
\end{align}

All terms involving \(\bm G_1\), \(\bm G_3\), \(\bm G_5\),
\(\bm G_6\), and \(\bm G_7\) cancel because the two paths are identical
outside the error interval. Using
\(\bm e=\bar{\bm a}_e-\bm a_e\),
\(\bm\Psi=\bm G_2\),
\(\bm\Phi=\bm G_4\), and
\(\bm\epsilon_f=\bar{\bm a}_f-\hat{\bm a}_f\), we obtain
\begin{align}
\bar{\mathcal M}-\mathcal M
=
\bm e^\dagger\bm\Psi\bm e
+
2\operatorname{Re}\{
\bm e^\dagger\bm\Phi\bm\epsilon_f
\}
+
2\operatorname{Re}\{
\bm e^\dagger\bm\omega
\}.
\label{eq:appendix_metric_difference}
\end{align}

An error event occurs when the competing path has a larger metric,
that is, when \(\mathcal M>\bar{\mathcal M}\). Therefore, an error
event occurs if
\begin{align}
\bm e^\dagger\bm\Psi\bm e
+
2\operatorname{Re}\{
\bm e^\dagger\bm\Phi\bm\epsilon_f
\}
+
2\operatorname{Re}\{
\bm e^\dagger\bm\omega
\}
<0 .
\label{eq:appendix_error_condition}
\end{align}

In the absence of bias correction,
\(\hat{\bm a}_f=\bm 0\), and therefore
\(\bm\epsilon_f=\bar{\bm a}_f\). In this case,
\eqref{eq:appendix_error_condition} reduces to
\[
\bm e^\dagger\bm\Psi\bm e
+
2\operatorname{Re}\{
\bm e^\dagger\bm\Phi\bar{\bm a}_f
\}
+
2\operatorname{Re}\{
\bm e^\dagger\bm\omega
\}
<0 .
\]
The discussions above show that the
metric comparison between the competing
and transmitted sequence is also affected by the correlation between the symbols from traceback and the undecided future symbols. This residual future-symbol
term can reduce the effective RKHS separation between the two paths and
increase the probability of an incorrect path decision. Bias correction
reduces this effect by replacing $\bar{\boldsymbol a}_f$ with the
future-symbol decision error
$\bm \epsilon_f=\bar{\boldsymbol a}_f-\hat{\boldsymbol a}_f$; as the
future-symbol decisions become more accurate, this term approaches zero
and the error event is determined increasingly by the RKHS distance and
the filtered-noise projection.

\section{Pseudocode for the algorithm described in this paper}

In this appendix, we provide the pseudocode for the
reduced-state Ungerboeck-type BCJR receiver for FTN described
in Section~\ref{sec:decodersecBredstate}.

\begin{algorithm}[htbp]
\small
\caption{Reduced-state Ungerboeck-type BCJR receiver with
bidirectional decision feedback for FTN}
\label{alg:ftn_rs_bcjr}
\begin{algorithmic}[1]

\Require
$\boldsymbol{y}$, $\mathcal{A}$ with $|\mathcal{A}|=M$, $N_0$,
full memory $L$, reduced memory $J\leq L$, symbol priors $p(a_k)$,
allowed boundary states $\sigma_0,\sigma_N$, and number of iterations
$I$.

\Ensure
$q_{J,k}(c;\boldsymbol{y})$ and $\hat a_k$ for
$k=0,\ldots,N-1$.

\State Set
\[
\sigma_k=(a_{k-1},\ldots,a_{k-J})
\]
and retain only transitions satisfying
$T(\sigma_k,a_k,\sigma_{k+1})=1$.

\State For every iteration, initialize the boundary messages as
\[
\ln\alpha^f_0(\sigma_0)=0,\qquad
\ln\alpha^b_N(\sigma_N)=0,
\]
with all other invalid boundary messages equal to $-\infty$.

\For{$\iota=1,\ldots,I$}

    \If{$\iota=1$}
        \State Set $B_k=0$ for all $k$.
    \Else
        \State Trace back the forward survivors to obtain
        $\{\dot a_i\}_{i=k-L+1}^{k-J}$
        and trace back the backward survivors to obtain
$\{\hat a_i\}_{i=k+1}^{k+L-J}$.
        \State Evaluate $B_k$ from \eqref{eqn:bias_definition}, using only valid indices
        at the block boundaries.
    \EndIf

    \Statex \textbf{Forward recursion and survivor construction}

    \For{$k=0,\ldots,N-1$}

        \For{each valid transition
        $\sigma_k\xrightarrow{a_k}\sigma_{k+1}$}

            \State Evaluate the reduced-state branch metric
            $\widetilde{J}(a_k,\sigma_k)$ from (24).

            \State Form the bias-corrected survivor metric using \eqref{eqn:logforwardmsgupdate}.
            % \[
            % \begin{aligned}
            % &\ln\alpha^f_k(\sigma_k)
            % +\ln\widetilde{J}(a_k,\sigma_k)
            % +\ln p(a_k)\\
            % &\qquad
            % -\frac{B_k}{N_0}.
            % \end{aligned}
            % \]

        \EndFor

        \State For each destination state $\sigma_{k+1}$, retain the
        predecessor and symbol that maximize the preceding expression.
        Store the corresponding survivor $a_k^\star,\sigma_k^\star$.

        \State Store the corresponding uncorrected forward metric:
        \[
        \begin{aligned}
        \ln\alpha^f_{k+1}(\sigma_{k+1})
        &=
        \ln\alpha^f_k(\sigma_k^\star)
        +\ln\widetilde{J}(a_k^\star,\sigma_k^\star)\\
        &\quad+\ln p(a_k^\star).
        \end{aligned}
        \]

    \EndFor

    \algstore{ftnreceiver}

\end{algorithmic}
\end{algorithm}

\begin{algorithm}[!t]
\small
\addtocounter{algorithm}{-1}
\caption{Reduced-state Ungerboeck-type BCJR receiver with
bidirectional decision feedback for FTN (continued)}
\begin{algorithmic}[1]

\algrestore{ftnreceiver}

\Statex \textbf{Backward recursion and survivor construction}

\For{$k=N-1,\ldots,0$}

    \State Recompute $B_k$ from (25), using the updated forward
    survivor and the current backward traceback.

    \For{each valid transition
    $\sigma_k\xrightarrow{a_k}\sigma_{k+1}$}

        \State Form the bias-corrected backward metric using \eqref{eqn:logbackwardmsgupdate}
        % \[
        % \begin{aligned}
        % &\ln\alpha^b_{k+1}(\sigma_{k+1})
        % +\ln\widetilde{J}(a_k,\sigma_k)
        % +\ln p(a_k)\\
        % &\qquad
        % -\frac{B_k}{N_0}.
        % \end{aligned}
        % \]

    \EndFor

    \State For each predecessor state $\sigma_k$, retain the outgoing
    transition that maximizes the preceding expression.
    Store the backward-survivor $a_k^\star,\sigma_{k+1}^\star$.

    \State Store the corresponding uncorrected backward metric:
    \[
    \begin{aligned}
    \ln\alpha^b_k(\sigma_k)
    &=
    \ln\alpha^b_{k+1}(\sigma_{k+1}^\star)
    +\ln\widetilde{J}(a_k^\star,\sigma_k)\\
    &\quad+\ln p(a_k^\star).
    \end{aligned}
    \]

\EndFor

\State Trace back the backward survivors to update the tentative
future decisions $\{\hat a_k\}$.

\EndFor

\Statex \textbf{Final symbol scores and decisions}

\For{$k=0,\ldots,N-1$}
    \For{each $c\in\mathcal{A}$}

        \State Compute the final reduced-state score using \eqref{eqn:posteriorandscorefull}.
        % \[
        % \begin{aligned}
        % \ln q_{J,k}(c;\boldsymbol{y})
        % &=
        % \ln p(c)\\
        % &\quad+
        % \max_{\substack{\sigma_k,\sigma_{k+1}:\\
        % T(\sigma_k,c,\sigma_{k+1})=1}}
        % \Bigg[
        % \ln\alpha^f_k(\sigma_k)
        % +\ln\widetilde{J}(c,\sigma_k)\\
        % &\qquad\qquad
        % +\ln\alpha^b_{k+1}(\sigma_{k+1})
        % -\frac{B_k}{N_0}
        % \Bigg].
        % \end{aligned}
        % \]

    \EndFor

    \State Decide
    \[
    \hat a_k
    =
    \arg\max_{c\in\mathcal{A}}
    q_{J,k}(c;\boldsymbol{y}).
    \]

\EndFor

\end{algorithmic}
\end{algorithm}

\bibliographystyle{IEEEtran}

\bibliography{main}

\end{document}